\documentclass[journal, twoside, final]{IEEEtran}
\usepackage{amsmath,amssymb,amsfonts}
\usepackage{algorithm}
\usepackage{array}
\usepackage{makecell}
\usepackage[caption=false,font=normalsize,labelfont={scriptsize,sf},textfont=sf]{subfig}
\usepackage{textcomp}
\usepackage{tabularx}
\usepackage{stfloats}
\usepackage{url}
\usepackage{graphicx}

\newcolumntype{C}[1]{>{\hsize=#1\hsize\linewidth=\hsize\centering\arraybackslash}X}
\usepackage{cite}
\usepackage{xcolor}
\usepackage{tikz}
\usetikzlibrary{arrows.meta,calc,positioning}
\usepackage{algpseudocode}
\usepackage{float}
\usepackage{placeins}
\usepackage{siunitx}
\usepackage[
  colorlinks=true,
  linkcolor=blue,
  citecolor=blue
]{hyperref}

\newtheorem{theorem}{Theorem}
\newtheorem{corollary}{Corollary}
\newtheorem{lemma}{Lemma}
\newtheorem{rmk}{Remark}
\newtheorem{definition}{Definition}

\begin{document}

\title{Unified Geometric--Kinematic Laws of Mobility in 3D Cellular Networks}
\author{Tian Shi and Minghua Xia,~\IEEEmembership{Senior Member,~IEEE}
 	\thanks{Manuscript received 12 April 2026; revised 2 August 2026; accepted 10 October 2026. This work was supported in part by the Postgraduate Innovation Program of Sun Yat-sen University under Grant 09010-12263017 and in part by the Guangdong Basic and Applied Basic Research Foundation under Grant 2025A1515140074. The associate editor coordinating the review of this article and approving it for publication was M. Kishk. \textit{(Corresponding author: Minghua Xia)}}
	\thanks{Tian Shi is with the School of Electronics and Information Technology, Sun Yat-sen University, Guangzhou 510006, China (email: shit26@mail2.sysu.edu.cn).}
	\thanks{Minghua Xia is with the School of Electronics and Information Technology, Sun Yat-sen University, Guangzhou 510006, China, and also with Southern Marine Science and Engineering Guangdong Laboratory, Zhuhai 519082, China (email: xiamingh@mail.sysu.edu.cn).} 
	\thanks{Digital Object Identifier 10.1109/TWC.2026.3743548}
}

\markboth{IEEE Transactions on Wireless Communications} {Shi \MakeLowercase{\textit{et al.}}: Unified Geometric--Kinematic Laws of Mobility in 3D Cellular Networks}

\maketitle

\IEEEpubid{\begin{minipage}{\textwidth} \ \\[12pt] \centering 1536-1276 \copyright\ 2026 IEEE. All rights reserved, including rights for text and data mining, and training of artificial intelligence \\ and similar technologies. Personal use is permitted, but republication/redistribution requires IEEE permission. \\
See \url{https://www.ieee.org/publications/rights/index.html} for more information.\end{minipage}}
	
\maketitle

\maketitle

\begin{abstract}
Mobility in three-dimensional (3D) cellular networks induces frequent handovers and random sojourn times for aerial users. Analyzing these events is challenging because they jointly depend on user trajectories and random service-boundary geometry. This paper develops a unified geometric--kinematic framework in which service regions form a stationary random tessellation, their boundaries constitute a random closed set, and user motion follows a cruise--turn mobility model. Under the stated stationarity, isotropy, and independence conditions, the handover rate satisfies the exact law \(H=(S_V/2)\bar v\), where \(S_V\) is the specific surface area and \(\bar v\) is the long-run mean speed. The forward boundary-distance distribution provides the local geometric input to an effective-traversal approximation for the mean remaining sojourn time. These results link global and local geometric descriptors to complementary mobility metrics. Combining the handover law with link coverage yields a handover-aware effective spectral efficiency for base station (BS) density selection and service-architecture comparison. We then specialize the framework to Poisson--Voronoi single-BS service and Poisson--Delaunay four-BS coordinated multi-point service; under both models, the handover rate scales as \(\lambda^{1/3}\bar v\). Numerical results corroborate the exact handover expressions, assess the sojourn-time approximations, and show how user speed and interruption duration affect the preferred density and architecture.
\end{abstract}


 \IEEEpubidadjcol
 
%%------------------------------------------------------------------------------------
\section{Introduction}
\label{sec:introduction}

\IEEEPARstart{R}{ecent} advances in uncrewed aerial vehicle (UAV) technology and space--air--ground integrated networks are accelerating the development of three-dimensional (3D) cellular architectures for aerial connectivity and low-altitude wireless networking~\cite{yuan2026}. By exploiting altitude and mobility, UAV communications can support emergency connectivity~\cite{Zhang2023}, large-scale data collection~\cite{9515586}, and terrestrial coverage enhancement~\cite{11311373}. These applications make reliable service continuity under aerial mobility increasingly important.

A defining feature of UAV cellular networks is the interaction between user mobility and genuinely 3D service geometry~\cite{9861699}. As an aerial user moves, crossing a service-domain boundary changes its serving entity and triggers a \emph{handover}. The time from a given observation instant to the next boundary crossing is the \emph{remaining sojourn time}. These quantities affect signaling overhead, service interruption, association stability, and ultimately the communication performance experienced under mobility~\cite{8673556}.

The principal analytical difficulty lies in coupling the user trajectory with the irregular boundaries of random service regions. The handover rate is a long-run boundary-crossing frequency, whereas the remaining sojourn time is a local first-exit quantity. Characterizing both therefore requires geometric descriptors that capture how a spatial trajectory intersects random boundary surfaces.

 \IEEEpubidadjcol
 
\subsection{Related Work and Research Gap}

Most stochastic-geometric mobility studies assume planar base station (BS) deployments. Trajectory-based methods treat handovers as path--boundary intersections and derive handover rates and sojourn times \cite{6477064}. Association-based methods track serving-BS changes and study coverage, mobility-aware association, handover skipping, and interference coordination~\cite{7110473,7006787,7827020,10130125,10427200,10464354}. These studies provide key foundations but retain two-dimensional service geometry.

For aerial users, the random walk, random waypoint, and random direction models provide useful analytical baselines \cite{camp2002survey,Bettstetter2004,8671460,10777167}. However, their conventional forms do not jointly capture heading evolution on the unit sphere, finite-duration reorientation, and platform-dependent translation during turning. Explicit aerial-user mobility has been considered for handover and coverage analysis under planar BS deployments~\cite{8998329,10005193}, while related studies have examined handover failure and ping-pong effects without specifying a complete temporal trajectory-evolution model~\cite{10125057}. In a genuinely 3D network, spatial trajectories intersect random boundary surfaces rather than planar boundary curves.
 
A parallel line of stochastic-geometric work has considered genuinely 3D BS deployments, analyzing distance distributions, latency, spectral efficiency, coverage probability, and achievable rate \cite{8533634,9151343,9740446}. These studies focus predominantly on static communication performance. The closest mobility-related study \cite{10345607} derives the expected number of handovers and the handover rate per unit travel distance in a 3D deployment, but does not incorporate an explicit temporal mobility process or characterize remaining sojourn time.

The present work builds on three established foundations: the trajectory--boundary interpretation of handovers, smooth-turn airborne mobility models, and known geometric descriptors of Poisson--Voronoi and Poisson--Delaunay tessellations. As summarized in Table~\ref{tab:literature_comparison}, previous studies have not jointly considered an explicit aerial-user trajectory, genuinely 3D service boundaries, and both temporal handover rate and remaining sojourn time. We address this gap by coupling an explicit 3D cruise--turn mobility process with a boundary representation of stationary service tessellations.

\begin{table}[!t]
\centering
\renewcommand{\arraystretch}{1.15}
\caption{Comparison with representative studies of cellular mobility and 3D network performance.}
\label{tab:literature_comparison}
{\scriptsize
\setlength{\tabcolsep}{2.2pt}
\begin{tabularx}{\columnwidth}{!{\vrule width1.2pt}C{0.85}!{\vrule width1.2pt}C{0.45}|C{0.60}|C{1.25}|C{1.85}!{\vrule width1.2pt}}
\noalign{\hrule height 1.2pt}
\textbf{Ref.} & \textbf{BS} & \textbf{User} & \textbf{Explicit mobility model} & \textbf{Research metrics} \\
\noalign{\hrule height 1.2pt}
\cite{6477064} & 2D & Ground & $\checkmark$ & Handover rate; sojourn time \\
\hline
\cite{7110473} & 2D & Ground & $\times$ & Horizontal/vertical handover rates; data rate \\
\hline
\cite{7006787,7827020} & 2D & Ground & $\times$ & Handover rate; coverage probability \\
\hline
\cite{8998329} & 2D & Air & $\checkmark$ & Handover rate; coverage probability \\
\hline
\cite{10005193} & 2D & Air & $\checkmark$ & Handover probability; coverage probability \\
\hline
\cite{10125057} & 2D & Air & $\times$ & Handover, failure, and ping-pong probabilities \\
\hline
\cite{8533634} & 3D & Air & $\times$ & Latency; spectral efficiency \\
\hline
\cite{9151343} & 3D & Air & $\times$ & Coverage probability; achievable rate \\
\hline
\cite{9740446} & 3D & Air & $\times$ & Distance distribution; SINR coverage probability \\
\hline
\cite{10345607} & 3D & Ground & $\times$ & Expected handovers; handover rate per unit length \\
\hline
\textbf{This work} & \textbf{3D} & \textbf{Air} & \textbf{$\checkmark$}
& \textbf{Handover rate; remaining sojourn time; effective spectral efficiency} \\
\noalign{\hrule height 1.2pt}
\end{tabularx}
\medskip
\par\noindent
\parbox{\columnwidth}{\scriptsize\emph{Note:} ``2D'' denotes fixed-height or height-marked planar BS deployments, whereas ``3D'' denotes BSs distributed in 3D space. A checkmark indicates an explicit trajectory-generation or state-evolution model for user mobility; a cross indicates otherwise.}
}
\end{table}

\subsection{Key Idea and Main Contributions}

We represent the union of service-domain boundaries as a boundary random closed set and interpret handovers as trajectory--boundary crossings. For stationary and isotropic service geometry independent of the mobility process, this representation yields an exact factorization of the handover rate into boundary density and long-run mean speed. By contrast, the mean remaining sojourn time is approximated using the forward boundary distance and the cruise--turn traversal dynamics. Because the formulation operates at the level of the induced service tessellation, it accommodates both single-BS association and coordinated multi-point (CoMP) service.

The main contributions are as follows.

\begin{enumerate}

\item \emph{Tessellation-level mobility formulation:} We formulate mobility-induced service transitions through intersections between a continuous 3D trajectory and the boundary of a stationary service tessellation. This separates the induced service geometry from the particular deployment, association, or cooperation mechanism used to construct it.

\item \emph{Exact handover law and 3D scaling:} For stationary and isotropic service boundaries and an independent mobility model with a stationary and ergodic velocity process, we derive the handover rate \( H=\frac{S_V}{2}\bar v, \) where \(S_V\) is the specific surface area and \(\bar v\) is the long-run mean speed. The cruise--turn model yields explicit platform-dependent expressions for \(\bar v\), while other stationary mobility models enter through their respective mean speeds. For the considered 3D Poisson models, \(H\propto\lambda^{1/3}\bar v\), in contrast to the \(H\propto\lambda^{1/2}\bar v\) scaling of planar Poisson deployments.

\item \emph{Remaining-sojourn-time approximation:} We use the forward boundary distance as the local geometric input to an effective-traversal approximation for the mean remaining sojourn time. This provides a complementary description of residence behavior while maintaining a clear distinction between the exact handover law and the approximate sojourn-time results.

\item \emph{Architecture specialization and network design:} We specialize the framework to Poisson--Voronoi nearest-BS service and Poisson--Delaunay four-BS CoMP service. A handover-aware effective spectral efficiency then combines coverage performance with mobility-induced interruption, enabling BS-density optimization and comparison of the two architectures under different mobility and interruption conditions.

\end{enumerate}

The remainder of this paper is organized as follows. Section~\ref{sec:system_model} presents the service geometry and mobility models. Section~\ref{sec:performance} derives the exact handover law, the approximate mean-sojourn-time expression, and the handover-aware performance metric. Section~\ref{sec:applications} specializes the analysis to Poisson--Voronoi and Poisson--Delaunay networks. Section~\ref{sec:numerical_results} presents the numerical results, and Section~\ref{sec:conclusion} concludes the paper.

\emph{Notation:} Scalars are denoted by italic letters, vectors by bold lowercase letters, and matrices by bold uppercase letters. The sets \(\mathbb R^3\) and \( \mathbb S^2\triangleq \{\mathbf u\in\mathbb R^3:\|\mathbf u\|=1\} \) denote 3D Euclidean space and the unit sphere, respectively. For \(\mathbf x,\mathbf y\in\mathbb R^3\), \(\|\mathbf x\|\) and \(\langle\mathbf x,\mathbf y\rangle\) denote the Euclidean norm and inner product. The symbols \(\mathbb E[\cdot]\) and \(\mathbb P(\cdot)\) denote expectation and probability. The notation \(|\cdot|\) denotes Lebesgue measure for spatial sets and cardinality for finite sets.

\section{System and Mobility Models}
\label{sec:system_model}

This section introduces the system and mobility models. We first represent the 3D service regions by a stationary tessellation and define their boundary random closed set. We then introduce a cruise--turn mobility model for multirotor and fixed-wing UAVs. These models provide the geometric and kinematic foundations for the subsequent analysis of handover rate and remaining sojourn time.

\subsection{System Model}
\label{subsec:network_model}

\begin{figure}[!t]
  \centering
  \includegraphics[width=\columnwidth]{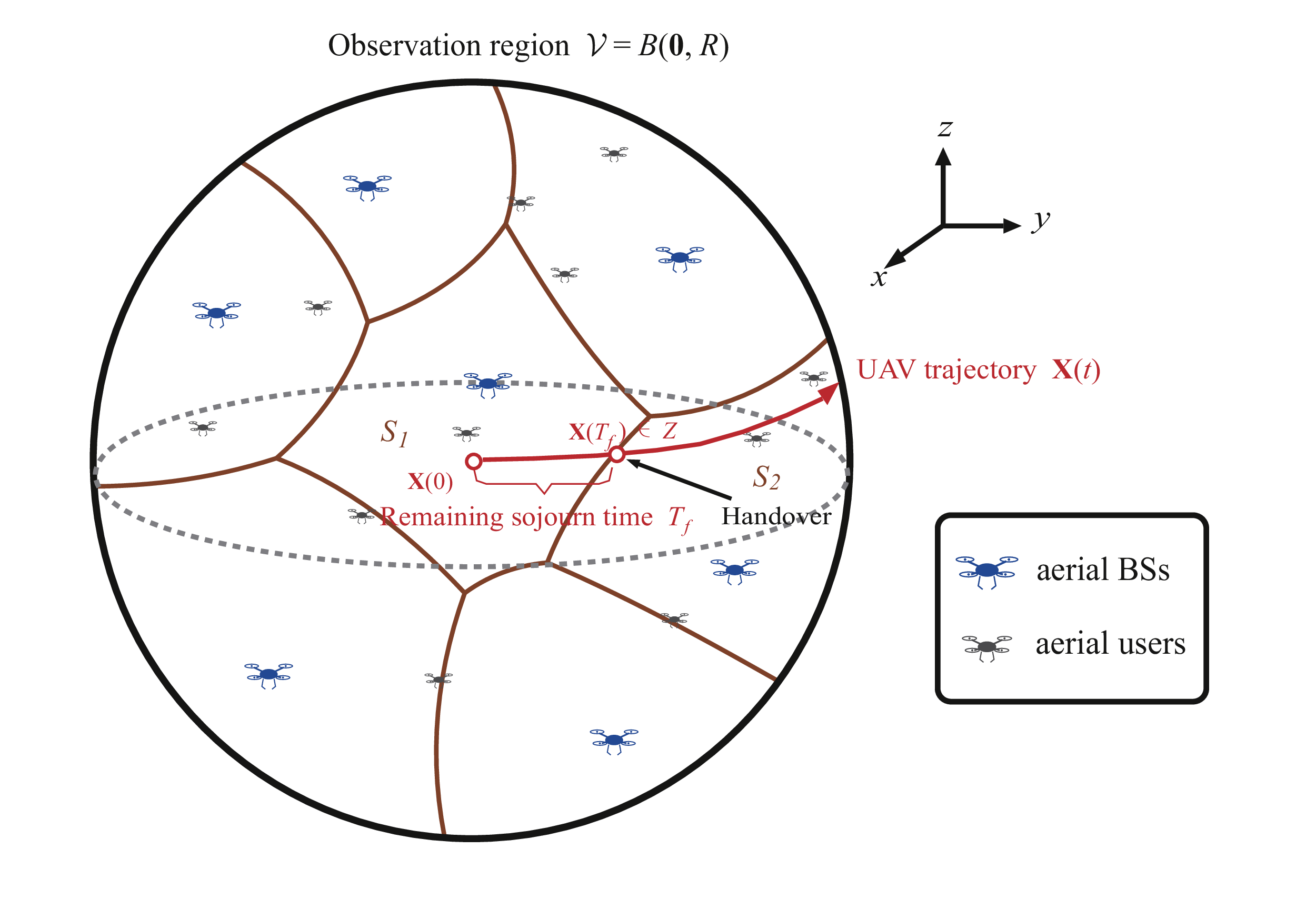}
\caption{System model for a 3D aerial network observed within a finite spherical region $\mathcal V=B(\mathbf 0,R)$. A homogeneous PPP $\Phi\subset\mathbb R^3$ models the aerial-BS locations, and only the BSs within $\mathcal V$ are displayed in the figure. The brown curves schematically illustrate representative service-region boundaries, and $\mathbf X(t)$ denotes an aerial-user trajectory. The elapsed time until the first boundary hit $\mathbf X(T_f)\in Z$ is the remaining sojourn time $T_f$; each service-boundary crossing triggers a handover or serving-set update.}
  \label{fig:system_model}
\end{figure}

Motivated by finite-domain 3D aerial-network models \cite{8533634,9151343,9454462}, we consider an aerial network observed within a spherical region $\mathcal V=B(\mathbf 0,R)\subset\mathbb R^3$. The aerial BS locations are modeled by a homogeneous Poisson point process (PPP) $\Phi\subset\mathbb R^3$ with intensity $\lambda>0$, and the BS configuration within the region is denoted by $\Phi_{\mathcal V}=\Phi\cap\mathcal V$. We restrict attention to trajectory segments lying in the central interior of $\mathcal V$, where the local service geometry is represented by the stationary full-space tessellation induced by $\Phi$. Consequently, the observation boundary $\partial\mathcal V$ is not considered in the subsequent service-boundary-crossing analysis. Fig.~\ref{fig:system_model} illustrates the spherical observation region, representative service-region boundaries, and an aerial-user trajectory.

The adopted service rule induces a stationary, isotropic, and face-to-face tessellation \(\mathcal C(\Phi,\mathcal M)\) of \(\mathbb R^3\), where \(\mathcal M\) is an auxiliary mark process representing service-dependent attributes. This construction includes Poisson--Voronoi and Poisson--Delaunay tessellations \cite{moller1989random} and extends to weighted models such as Laguerre \cite{Lautensack_2008} and Johnson--Mehl tessellations~\cite{moller_1992}. The cells of \(\mathcal C(\Phi,\mathcal M)\) have disjoint interiors and cover \(\mathbb R^3\). For any location \(\mathbf x\) outside the cell boundaries, let \(C(\mathbf x)\) denote the unique cell containing \(\mathbf x\) in its interior, and let \(\mathcal S(\mathbf x)\subseteq\Phi\) be the serving BS set specified by the service rule.

\begin{definition}[Boundary random closed set]
\label{def:boundary_rcs}
The service-boundary random closed set is
\begin{equation}
\label{eq:boundary}
Z\triangleq
\bigcup_{C\in\mathcal C(\Phi,\mathcal M)}\partial C,
\end{equation}
where \(\partial C\) denotes the boundary of cell \(C\).
\end{definition}

For a regular bounded observation window \(W\subset\mathbb R^3\), define the specific surface area of \(Z\) by
\begin{equation}
S_V
\triangleq
\lim_{r\to\infty}
\frac{\mathbb E\!\left[\operatorname{Area}(Z\cap rW)\right]}
{|rW|},
\end{equation}
provided that the limit exists and is finite. The quantity \(S_V\), measured in \(\si{m^{-1}}\), is the mean boundary area per unit volume.

The formulation operates at the level of the service tessellation induced by the adopted rule. The mark process \(\mathcal M\) need not appear explicitly in the final mobility expressions because its effects are captured by the boundary descriptors of the induced tessellation, principally \(S_V\) and the forward boundary-distance distribution. Poisson--Voronoi and Poisson--Delaunay tessellations are considered in Subsections~\ref{subsec:pv_application} and~\ref{subsec:pd_application} as tractable examples of single-BS and cooperative service architectures, respectively.

\begin{rmk}[Scope of Deployment and Antenna Assumptions]
The homogeneous PPP and the resulting Poisson--Voronoi and Poisson--Delaunay models are analytical benchmarks rather than empirically validated descriptions of terrestrial-network reuse or dedicated low-altitude aerial deployments. Collision avoidance, minimum-separation constraints, corridors, and planned layouts may instead induce repulsion or spatial inhomogeneity. The general mobility analysis remains applicable when the induced service tessellation satisfies the stated conditions and its boundary descriptors are available; for repulsive deployments, these descriptors typically require numerical evaluation.

Spatial isotropy refers to rotational invariance of the service-boundary statistics and does not require an isotropic antenna radiation pattern. Antenna patterns affect the desired and interfering link gains and, under gain-based association, may also alter the service regions. Directional or elevation-dependent patterns can be incorporated through the mark process and the induced association rule. If the resulting service geometry is anisotropic, the scalar factor \(S_V/2\) must be replaced by a direction-dependent boundary-intersection intensity, and the corresponding directional forward boundary-distance distribution must be used for sojourn-time analysis.
\end{rmk}

Having specified the service geometry, we next introduce the mobility model. The interaction between the UAV trajectory and the service boundaries determines the handover and remaining-sojourn behavior analyzed below.

\subsection{Mobility Model}
\label{subsec:mobility_model}

Consider the trajectory of an aerial user (e.g., a UAV) $\{\mathbf X(t)\}_{t\ge 0}$ evolving in $\mathbb{R}^3$, governed by
\begin{equation}
\dot{\mathbf X}(t)=V(t)\mathbf U(t),
\end{equation}
where $\mathbf U(t)\in\mathbb{S}^2$ denotes the unit flight direction and $V(t)\ge 0$ is the translational speed magnitude.

The UAV motion alternates between \emph{cruise} and \emph{turn} phases, capturing intermittent changes in flight direction. This construction builds on smooth-turn airborne mobility models that preserve directional persistence and finite-rate reorientation~\cite{6476039,8325272}. Exponential cruise durations provide a memoryless renewal structure~\cite{6476039}, while the turn-phase speed \(v_t\) distinguishes platform-dependent translational behavior during reorientation.

\subsubsection*{1) Cruise Phase}

During the $k$-th cruise phase, the UAV moves at constant speed along a fixed direction:
\begin{equation}
V(t)=v_c, \qquad \mathbf U(t)=\mathbf U_k,
\end{equation}
where $v_c>0$ is the cruising speed and $\mathbf U_k\in\mathbb{S}^2$ is the heading direction.

The cruise duration follows an exponential distribution:
\begin{equation}
T_{c,k}\sim \mathrm{Exp}(\nu),
\end{equation}
where $\nu>0$ is the reorientation triggering rate, measured in \si{s^{-1}}; hence, $1/\nu$ is the mean cruise duration. If the $k$-th cycle starts at position $\mathbf X_k$, then the position at the end of the cruise phase is
\begin{equation}
\mathbf X_k^{\mathrm{c}} = \mathbf X_k + v_c T_{c,k} \mathbf U_k.
\end{equation}

\subsubsection*{2) Turn Phase}

At the end of the cruise phase, the UAV updates its heading. A new direction $\mathbf U_{k+1}$ is sampled isotropically on the unit 3D sphere:
\begin{equation}
\mathbf U_{k+1}\sim \mathrm{Unif}(\mathbb{S}^2),
\end{equation}
independently of the past trajectory.

The turning angle is
\begin{equation}
\Theta_k = \arccos\!\big(\langle \mathbf U_k,\mathbf U_{k+1}\rangle\big),
\qquad \Theta_k\in[0,\pi].
\end{equation}

The reorientation follows the shortest great-circle arc with constant angular speed $\omega_{\max}>0$ (in \si{rad/s}), yielding the turn duration
\begin{equation}
\tau_k = \frac{\Theta_k}{\omega_{\max}}.
\end{equation}

For $\Theta_k\in(0,\pi)$, define the spherical linear interpolation (SLERP) \cite{shoemake1985animating}:
\begin{align*}
\mathrm{Slerp}(\mathbf U_k,\mathbf U_{k+1};s)
&= \frac{\sin((1-s)\Theta_k)}{\sin\Theta_k}\mathbf U_k + \frac{\sin(s\Theta_k)}{\sin\Theta_k}\mathbf U_{k+1}, \\
&\qquad s\in[0,1].
\end{align*}

Thus, during the turn phase,
\begin{equation}
\mathbf U(t) = \mathrm{Slerp}\!\left(\mathbf U_k,\mathbf U_{k+1};\frac{t}{\tau_k}\right),
\qquad 0\le t\le \tau_k .
\end{equation}
If $\Theta_k=0$, then no reorientation is needed and $\mathbf U(t)\equiv \mathbf U_k$.

During turning, the translational speed is constant:
\begin{equation}
V(t)=v_t, \qquad v_t\in[0,v_c].
\end{equation}
Accordingly, the position trajectory evolves as
\begin{equation}
\mathbf X_k^{\mathrm{turn}}(t)
= \mathbf X_k^{\mathrm c} + v_t\int_0^t \mathrm{Slerp}\!\left(\mathbf U_k,\mathbf U_{k+1};\frac{s}{\tau_k}\right)\,\mathrm{d}s,
\end{equation}
where $0 \le t\le \tau_k$. The next cycle starts from
\begin{equation}
\mathbf X_{k+1} = \mathbf X_k^{\mathrm{turn}}(\tau_k).
\end{equation}
Here \(k\) indexes a complete cruise--turn cycle, not a unit-time interval: cycle \(k+1\) begins only after the turn of duration \(\tau_k\) is completed. A terminal partial cycle contributes only a boundary term to the long-run rate.

\subsubsection*{3) Canonical UAV Mobility Regimes}

Two canonical UAV mobility regimes are considered (see Fig.~\ref{fig:UAV_trajectories}), distinguished by the translational behavior during the turning phase:

\begin{figure}[!t]
\centering
\captionsetup[subfigure]{margin=1pt}
\subfloat[\scriptsize Multirotor pause-to-turn trajectory.]{%
\includegraphics[width=3in]{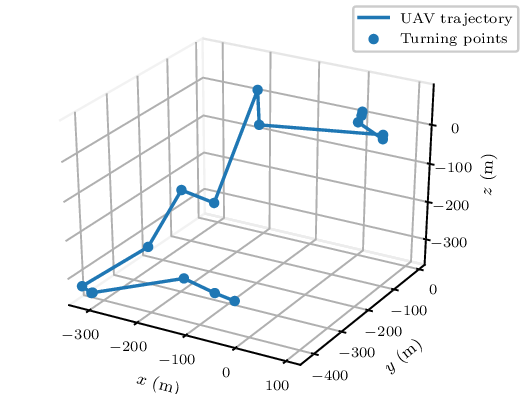}}%
\\
\subfloat[\scriptsize Fixed-wing fly-to-turn trajectory.]{%
\includegraphics[width=3in]{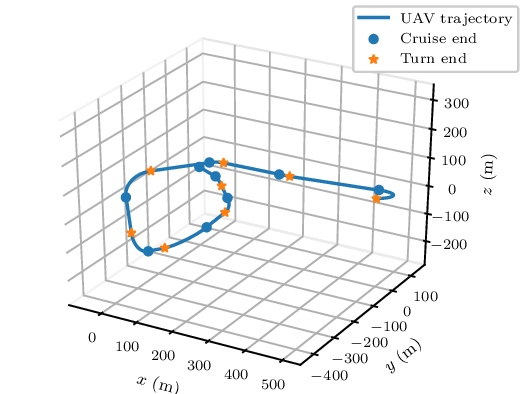}}
\caption{Example 3D UAV trajectories under the two canonical cruise--turn mobility regimes: (a) multirotor pause-to-turn motion and (b) fixed-wing fly-to-turn motion.}
\label{fig:UAV_trajectories}
\end{figure}

\begin{itemize}
\item \textit{Multirotor UAV (MR) (pause-to-turn):}
\begin{equation}
v_t=0,
\end{equation}
i.e., the UAV suspends its translation during reorientation, resulting in purely rotational motion during the turning phase. Consequently,
\begin{equation}
\mathbf X_{k+1}=\mathbf X_k^{\mathrm{c}}.
\end{equation}

\item \textit{Fixed-wing UAV (FW) (fly-to-turn):}
\begin{equation}
v_t=v_c,
\end{equation}
i.e., the UAV performs the turn while maintaining its cruising speed, resulting in continuous translational motion along a curved trajectory. The position update is given by
\begin{equation}
\mathbf X_{k+1}
= \mathbf X_k^{\mathrm{c}} + v_c\int_0^{\tau_k} \mathrm{Slerp}\!\left(\mathbf U_k,\mathbf U_{k+1};\frac{s}{\tau_k}\right)\,\mathrm{d}s .
\end{equation}
\end{itemize}

Since each post-turn direction is sampled uniformly and independently, the resulting mobility process is statistically isotropic, i.e., rotation-invariant in $\mathbb{R}^3$.

Notice that these regimes are idealized platform-dependent limits: multirotor hover supports pause-to-turn behavior~\cite{Mellinger2011}, whereas coordinated fixed-wing turning motivates fly-to-turn motion~\cite{GarciaMarina2017}. Intermediate behavior is represented by \(0<v_t<v_c\), and the parameter tuple \((v_c,v_t,\nu,\omega_{\max})\) can be calibrated to a given platform and mission profile.

\begin{rmk}[Mobility Scope and Baselines]
The model describes stationary en-route cruise--turn motion and excludes takeoff, climb, descent, landing, obstacle avoidance, and waypoint- or handover-aware trajectory planning. Such nonstationary or network-coupled phases require phase-specific models; the present analysis isolates mobility-induced boundary crossings in the en-route regime.

The general handover law is not tied to the cruise--turn process: other stationary mobility models enter through their long-run mean speed. Constant-speed random-direction or random-walk motion is recovered using its corresponding mean speed, with instantaneous reorientation obtained as \(\omega_{\max}\to\infty\). A random-waypoint model can be treated similarly whenever a stationary regime exists.
\end{rmk}

%%------------------------------------------------------------------------------------
\section{Mobility-Induced Performance Analysis} \label{sec:performance}

This section develops the mobility-induced performance analysis. We first define the handover rate and remaining sojourn time. We then derive an exact handover law, develop an effective-traversal approximation for the mean remaining sojourn time, and combine the handover rate with link coverage through a handover-aware effective spectral efficiency.

\subsection{Performance Metrics} \label{subsec:metrics}
The performance of mobile users in a cellular network is fundamentally affected by the frequency with which the serving BS set changes and the duration a user remains within a given service region. From a long-term perspective, these effects are naturally quantified by rates measured over extended time horizons. We therefore begin by defining the handover rate, which captures the asymptotic frequency of service-region transitions induced by user mobility.

\begin{definition}[Handover Rate]
\label{def:handover_rate}
A \emph{handover} occurs when the UAV switches its service region, equivalently when $\mathbf X(t)$ crosses $Z$ and enters a neighboring cell through a shared facet. Let $N_{\mathrm{HO}}(T)$ denote the number of handovers in $[0,T]$. The handover rate is
\begin{equation}
H \triangleq \lim_{T\to\infty} \frac{\mathbb{E}\!\left[N_{\mathrm{HO}}(T)\right]}{T},
\end{equation}
provided the limit exists.
\end{definition}

While the handover rate captures the long-term frequency of service-region transitions, the remaining sojourn time reflects the local temporal stability experienced by a user from a typical observation point within a cell. These two metrics provide complementary perspectives on mobility-induced dynamics.

To formalize this notion, we introduce the remaining sojourn time, defined as the time until the next boundary crossing from a typical interior point.

\begin{definition}[Remaining Sojourn Time]
\label{def:sojourn}
Under the stationary regime, the remaining sojourn time is defined as the time from $t=0$ to the next boundary hit:
\begin{equation}
T_f \triangleq \inf\{t>0: \mathbf X(t)\in Z\},
\end{equation}
conditioned on $\mathbf X(0)\notin Z$ (which holds a.s.).
\end{definition}

The above definition corresponds to the forward recurrence time of the cell residence process under stationarity. From a stochastic-geometry perspective, it can be interpreted as the sojourn time observed from a typical interior point, which is consistent with the Palm viewpoint commonly adopted in spatial network analysis.

\subsection{Geometric Preliminaries} \label{subsec:geometric_prelim}
Before proceeding to quantitative analysis, we first establish a geometric regularity property of boundary-intersection events. This ensures that handover events can be characterized cleanly in terms of trajectory--boundary intersections.

\begin{lemma}
\label{lem:degenerate_hits}
Let $\{\mathbf X(t)\}_{t\ge 0}$ be an absolutely continuous trajectory, observed in stationarity from a typical interior point, and assume that the law of $\mathbf X(0)$ is absolutely continuous with respect to the 3D Lebesgue measure. Assume further that the motion randomness is non-degenerate in the sense that each new direction has a density on $\mathbb{S}^2$ and is independent of $\mathcal{C}$. Then, almost surely, each boundary intersection is transverse and occurs in the relative interior of a 2D facet of the tessellation, whereas intersections with edges or vertices and tangential facet contacts occur with probability zero. Consequently, the handover count equals the number of transverse intersections of $\mathbf X(t)$ with the boundary set $Z$, counted with multiplicity along the trajectory.
\end{lemma}

\begin{IEEEproof}
See Appendix~\ref{sec:degenerate_hits_proof}.
\end{IEEEproof}

\subsection{Handover Rate Analysis}
\label{sec:handover_rate_analysis}

With Lemma~\ref{lem:degenerate_hits}, handover events can be identified one-to-one with intersections between the UAV trajectory and the boundary set. This reduces the handover-rate analysis to counting boundary intersections per unit time, which can be addressed using tools from stochastic and integral geometry.

We now present the main result.

\begin{theorem}[Universal Handover Law]
\label{thm:handover}
Consider a stationary and isotropic random tessellation in \(\mathbb R^3\) with a boundary random closed set \(Z\) of finite specific surface area \(S_V\). Let \(\{\mathbf X(t)\}_{t\geq0}\) be an independent mobility process with almost surely absolutely continuous sample paths and a stationary and ergodic velocity process. Assume further that its boundary contacts satisfy the nondegeneracy conditions of Lemma~\ref{lem:degenerate_hits}. Define its long-run mean speed by
\begin{equation}
\bar{v} \triangleq \lim_{T\to\infty}\frac{1}{T}\int_0^T V(t)\,\mathrm{d}t,
\end{equation}
provided the limit exists almost surely and is finite. Under ergodicity, $\bar v$ is almost surely constant.

Then, the handover rate admits the universal form
\begin{equation}
H = \frac{S_V}{2}\,\bar{v}.
\label{eq:handover_rate_main}
\end{equation}
\end{theorem}

\begin{IEEEproof}
See Appendix~\ref{sec:proof_handover}.
\end{IEEEproof}

Equation~\eqref{eq:handover_rate_main} separates geometry from mobility. The factor \(S_V/2\) is the mean number of boundary crossings per unit path length, whereas \(\bar v\) is the mean path length traveled per unit time. Their product therefore gives the mean number of handovers per unit time.

\begin{rmk}[Scope of the Universal Handover Law]
The term ``universal'' refers to the class specified in Theorem~\ref{thm:handover}: within this class, \eqref{eq:handover_rate_main} depends on the service geometry and mobility process only through \(S_V\) and \(\bar v\), rather than their detailed distributions. Stationarity removes location dependence, isotropy removes directional dependence and yields the factor \(S_V/2\), and independence excludes preferential coupling between the trajectory and service boundaries. In inhomogeneous, anisotropic, or trajectory-aware networks, the crossing intensity may depend on location, direction, or both; the scalar \(S_V/2\) must then be replaced by the corresponding local or directional boundary-intersection intensity.
\end{rmk}

To make Theorem~\ref{thm:handover} explicit under the cruise--turn mobility model, we next characterize the long-run mean speed.

\begin{lemma}[Long-Run Mean Speed]
\label{lem:vbar}
Under the cruise--turn mobility model with deterministic turn-phase translational speed $v_t\in[0,v_c]$, cruise duration $T_c\sim\mathrm{Exp}(\nu)$, and isotropic reorientation, the long-run mean speed $\bar v$ is given almost surely by
\begin{equation} \label{eq:MeanSpeed}
\bar v
= \frac{\mathbb{E}[v_c T_c]+\mathbb{E}[v_t\tau]} {\mathbb{E}[T_c]+\mathbb{E}[\tau]}
= \frac{\frac{v_c}{\nu}+v_t\,\mathbb{E}[\tau]} {\frac{1}{\nu}+\mathbb{E}[\tau]} .
\end{equation}
Moreover, if the reorientation directions $\mathbf U_k$ and $\mathbf U_{k+1}$ are independent and uniformly distributed on $\mathbb{S}^2$, then
\begin{equation}
\mathbb{E}[\Theta]=\frac{\pi}{2},
\qquad
\mathbb{E}[\tau]=\frac{\pi}{2\omega_{\max}}.
\end{equation}
\end{lemma}

\begin{IEEEproof}
See Appendix~\ref{sec:vbar_proof}.
\end{IEEEproof}

The expression in \eqref{eq:MeanSpeed} quantifies the effect of mobility parameters on the long-run mean speed. In particular, the reorientation triggering rate $\nu$ and turning speed $v_t$ jointly determine the fraction of time spent in turning versus cruising, thereby influencing the long-run mean speed and, consequently, the handover rate.

By combining Theorem~\ref{thm:handover} with Lemma~\ref{lem:vbar}, we obtain explicit expressions for the handover rate.

\begin{corollary}[Explicit Handover-Rate Expressions]
\label{cor:handover_explicit}
Under the assumptions of Theorem~\ref{thm:handover} and Lemma~\ref{lem:vbar}, the handover rate is
\begin{equation} \label{eq:H}
H = \frac{S_V}{2}\,
\frac{\frac{v_c}{\nu} + v_t\,\frac{\pi}{2\omega_{\max}}}
     {\frac{1}{\nu} + \frac{\pi}{2\omega_{\max}}}.
\end{equation}
In particular, \eqref{eq:H} reduces to
\begin{itemize}
\item \emph{Multirotor UAV ($v_t=0$):}
\begin{equation} \label{eq:H_MR}
H_{\mathrm{MR}} = \frac{S_V}{2}\,\frac{v_c}{1+\nu\frac{\pi}{2\omega_{\max}}}.
\end{equation}

\item \emph{Fixed-wing UAV ($v_t=v_c$):}
\begin{equation} \label{eq:H_FW}
H_{\mathrm{FW}}  = \frac{S_V}{2}\,v_c.
\end{equation}
\end{itemize}
\end{corollary}

The resulting expressions highlight a clear distinction between UAV platforms. For multirotor UAVs, \eqref{eq:H_MR} implies that increasing \(\nu\) increases the fraction of time spent paused during reorientation, thereby reducing the long-run mean speed and the handover rate. In contrast, \eqref{eq:H_FW} shows that, among the mobility parameters, the fixed-wing handover rate depends only on \(v_c\) and is independent of \(\nu\), reflecting continuous translation during turning. The general expression \eqref{eq:H} interpolates between these two limiting regimes for \(0<v_t<v_c\).

\subsection{Sojourn Time Analysis}
\label{sec:sojourn_analysis}

The handover rate depends on the global boundary descriptor \(S_V\), whereas the remaining sojourn time is a local first-exit quantity. Under the effective traversal approximation, its geometric input is the distance from the current location to the first boundary hit along the initial direction. The linear contact distribution is introduced because it is precisely the distribution of this forward boundary distance.

\begin{definition}[Linear Contact Distribution]
\label{def:lcd_forward}
Let \(Z\subset\mathbb R^3\) be a random closed set. Under the typical interior-point convention, condition on \(\mathbf 0\notin Z\). Let \(\mathbf U\sim\mathrm{Unif}(\mathbb S^2)\) be an isotropic random direction independent of \(Z\), and define the directed line segment
\begin{equation*}
\ell(r)\triangleq\{t\mathbf U:0\leq t\leq r\},
\qquad r\geq0.
\end{equation*}
The \emph{linear contact distribution} (LCD) of \(Z\) is
\begin{equation*}
H_{\ell}(r)
\triangleq
\mathbb P\!\left(
Z\cap\ell(r)\neq\varnothing
\,\middle|\,
\mathbf 0\notin Z
\right).
\end{equation*}
Equivalently, define the \emph{forward boundary distance} along \(\mathbf U\) by
\begin{equation*}
D\triangleq\inf\{s>0:s\mathbf U\in Z\}.
\end{equation*}
Thus, \(H_{\ell}\) is the cumulative distribution function of \(D\):
\begin{equation*}
H_{\ell}(r)
=
\mathbb P\!\left(D\leq r\mid\mathbf 0\notin Z\right),
\qquad r\geq0.
\end{equation*}
\end{definition}

When \(D\) has an absolutely continuous distribution, denote its probability density function (PDF) by \(f_D\). Under the effective traversal approximation, the mean remaining sojourn time is obtained by averaging the time required to traverse a forward boundary distance \(d\) against \(f_D(d)\).

We are now ready to characterize the mean sojourn time. For notational convenience, define the spatial reorientation intensity
\begin{equation*}
\kappa\triangleq\frac{\nu}{v_c},
\end{equation*}
measured in $\si{m^{-1}}$, and the normalized turn speed
\begin{equation*}
\zeta\triangleq\frac{v_t}{v_c}\in[0,1].
\end{equation*}

\begin{theorem}[Approximate Mean Sojourn-Time Law]
\label{thm:SojournTime}
Let $D\ge 0$ denote the forward distance to the first geometric intersection with the cell boundary along the initial direction of motion, with finite mean $\mathbb E[D]<\infty$. Under the cruise--turn mobility model, and under the effective traversal approximation that turning contributes a fractional time overhead without altering the geometric forward boundary distance $D$, the mean sojourn time is approximated by
\begin{align}
\label{eq:meanTf_general}
\mathbb E[T_f]
&\approx
\left(
\frac{1}{v_c}
+
\frac{\pi\kappa(1-\zeta)}{2\omega_{\max}}
\right)\mathbb E[D] \\
&=
\int_0^\infty
\left(
\frac{1}{v_c}
+
\frac{\pi\kappa(1-\zeta)}{2\omega_{\max}}
\right)
d\,f_D(d)\,\mathrm{d}d.
\end{align}
\end{theorem}

\begin{IEEEproof}
See Appendix~\ref{sec:proof_SojournTime}.
\end{IEEEproof}

The approximate sojourn-time expression in \eqref{eq:meanTf_general} complements the exact handover law. Whereas the handover rate depends on the global boundary density \(S_V\), the approximate mean remaining sojourn time depends on the local geometric descriptor \(\mathbb E[D]\), equivalently the first moment of \(f_D\). The mobility process contributes through an effective traversal factor accounting for both cruising and turning dynamics.

As a direct consequence, we obtain simplified expressions for canonical UAV regimes.

\begin{corollary}[Canonical UAV Mobility Regimes]
\label{cor:canonical}
Under the effective traversal approximation of Theorem~\ref{thm:SojournTime}, the general result in~\eqref{eq:meanTf_general} reduces to the following canonical cases:
\begin{itemize}
\item \emph{Multirotor UAV (pause-to-turn):} $v_t=0$ (i.e., $\zeta=0$), for which
\begin{equation} \label{eq:meanTf_MR}
\mathbb E[T_f] \approx \int_0^\infty\left(\frac{1}{v_c} + \frac{\pi\kappa}{2\omega_{\max}}\right)d\, f_D(d)\,\mathrm{d}d.
\end{equation}

\item \emph{Fixed-wing UAV (fly-to-turn):} $v_t=v_c$ (i.e., $\zeta=1$), for which
\begin{equation}  \label{eq:meanTf_FW}
\mathbb E[T_f] \approx \int_0^\infty \frac{1}{v_c}d\, f_D(d) \,\mathrm{d}d.
\end{equation}
\end{itemize}
\end{corollary}

For multirotor UAVs, \eqref{eq:meanTf_MR} accounts for the additional temporal overhead caused by pause-to-turn behavior. For fixed-wing UAVs, \eqref{eq:meanTf_FW} assigns no turn-duration overhead because translation continues at speed $v_c$. Both conclusions are understood within the effective traversal approximation, which retains the initial forward boundary distance despite subsequent reorientations.

\subsection{Handover-Aware Effective Spectral Efficiency}
\label{subsec:effective_spectral_efficiency}

To connect the handover law to communication-oriented network design, consider fixed-rate transmission with target signal-to-interference-plus-noise ratio (SINR) threshold \(\theta\). A successful transmission provides the spectral efficiency \(R_\theta=\log_2(1+\theta)\). For a service architecture \(\mathcal T\), let \(P_{\mathrm{cov}}^{\mathcal T}(\theta,\lambda)\) denote the probability that the instantaneous SINR exceeds \(\theta\), and let \(H^{\mathcal T}(\lambda,\bar v)\) denote the handover or serving-set-update rate.

Suppose that each handover or serving-set update causes a service interruption with mean duration \(\tau_a\). When interruptions are short relative to the mean inter-handover time and rarely overlap, the unavailable-time fraction is approximated by \(\tau_aH^{\mathcal T}(\lambda,\bar v)\). Following the separated-factor approach used in tractable mobility-aware analyses \cite{7006787,7827020,10130125,10427200,10720695,11126168}, we define the handover-aware effective spectral efficiency as
\begin{equation}
\label{eq:Ga_general}
G_a^{\mathcal T}(\lambda,\bar v;\theta,\tau_a)
\triangleq
\left[1-\tau_aH^{\mathcal T}(\lambda,\bar v)\right]^+
P_{\mathrm{cov}}^{\mathcal T}(\theta,\lambda)
\log_2(1+\theta),
\end{equation}
where \([x]^+\triangleq\max\{x,0\}\). The three factors in \eqref{eq:Ga_general} represent the approximate service-availability fraction, the link-success probability, and the fixed transmission rate, respectively. Thus, \(G_a^{\mathcal T}\) is measured in bit/s/Hz; multiplying it by the allocated bandwidth gives the corresponding effective data rate.

Equation~\eqref{eq:Ga_general} is a tractable performance metric rather than a joint temporal characterization of handover events and SINR. It separates mobility-induced service availability from link coverage and is most appropriate in the regime \(\tau_aH^{\mathcal T}(\lambda,\bar v)\ll1\). This construction exposes the tradeoff between the communication benefit of densification and its associated handover cost.

For prescribed \(\bar v\), \(\theta\), and \(\tau_a\), an architecture-dependent optimal BS density may be selected according to
\begin{equation}
\label{eq:lambda_star_general}
\lambda_{\mathcal T}^{\star}(\bar v;\theta,\tau_a)
\in
\underset{\lambda\in\Lambda}{\arg\max}\;
G_a^{\mathcal T}(\lambda,\bar v;\theta,\tau_a),
\end{equation}
where \(\Lambda\) is the admissible density set. For the 3D Poisson models considered below, \(H^{\mathcal T}(\lambda,\bar v)\propto\lambda^{1/3}\bar v\). Consequently, when the coverage probability initially increases and subsequently saturates with densification, the growing handover penalty can produce an interior optimal density.

\section{Specialization to Representative 3D Cellular Networks}
\label{sec:applications}

The general laws established in Section~\ref{sec:performance} are specialized to representative stochastic tessellation models, demonstrating how global and local geometric descriptors translate into explicit mobility performance metrics. Specifically, we consider two canonical 3D cellular architectures: Poisson--Voronoi (PV) networks under single-BS association and Poisson--Delaunay (PD) networks under CoMP-enabled transmission. These case studies provide instantiations of the unified geometric--kinematic framework and yield explicit performance expressions under tractable stochastic models.

\subsection{Poisson--Voronoi Cellular Networks}
\label{subsec:pv_application}

\subsubsection{Geometry and Association Model}

We first consider the case where the service regions are induced by a 3D Poisson--Voronoi tessellation generated by the BS point process $\Phi$. This model corresponds to the classical nearest-BS association rule widely adopted in cellular networks \cite{8533634}.

Given a locally finite point set $\Phi=\{\mathbf y_i\}\subset\mathbb{R}^3$, the Voronoi cell associated with $\mathbf y_i$ is defined as
\begin{equation}
\mathcal{V}(\mathbf{y}_i)
= \bigl\{\mathbf{x} \in \mathbb{R}^3 : \|\mathbf{x} - \mathbf{y}_i\| \le \|\mathbf{x} - \mathbf{y}_j\|, \ \forall\, \mathbf{y}_j \in \Phi \bigr\}.
\end{equation}

When $\Phi$ is a homogeneous PPP of intensity $\lambda$, the resulting tessellation is a Poisson--Voronoi tessellation. In this setting, each BS serves the users located within its Voronoi cell:
\begin{equation}
\mathcal S(\mathbf x) = \{\mathbf y_i\in\Phi: \mathbf x\in\operatorname{int}\mathcal V(\mathbf y_i)\}, \qquad |\mathcal S(\mathbf x)|=1.
\end{equation}
Thus, the general boundary random closed set reduces to the Voronoi cell boundaries, enabling direct application of the results in Section~\ref{sec:performance}.

\subsubsection{Specialization of the Handover Law}

We now specialize the universal handover law to the Poisson--Voronoi tessellation by substituting the corresponding specific surface area.

\begin{corollary}[Handover Rate in 3D Poisson--Voronoi Networks]
\label{cor:HO_PV_3D}
Consider a 3D Poisson--Voronoi tessellation generated by a homogeneous PPP of intensity $\lambda$, whose specific surface area is given by \cite[Eq. (9.104)]{chiu2013stochastic}
\begin{equation}
\label{eq:SV_PV_3D}
S_V^{\mathrm{PV}}
= 2\left(\frac{4\pi}{3}\right)^{1/3}\, \Gamma\!\left(\frac{5}{3}\right)\, \lambda^{1/3}.
\end{equation}
Under the general cruise--turn mobility model, the handover rate is given by
\begin{equation}
\label{eq:H_PV_general}
H
= \frac{S_V^{\mathrm{PV}}}{2}\,\frac{\frac{v_c}{\nu} + v_t \dfrac{\pi}{2\omega_{\max}}}{\dfrac{1}{\nu} + \dfrac{\pi}{2\omega_{\max}}}.
\end{equation}
In particular,
\begin{itemize}
\item Multirotor UAVs (pause-to-turn, $v_t=0$):
\begin{equation}
\label{eq:H_PV_MR}
H_{\mathrm{MR}}^{\mathrm{PV}}
= \left(\frac{4\pi}{3}\right)^{1/3}\,\Gamma\!\left(\frac{5}{3}\right)\,\lambda^{1/3}\frac{v_c}{1+\nu\dfrac{\pi}{2\omega_{\max}}}.
\end{equation}

\item Fixed-wing UAVs (fly-to-turn, $v_t=v_c$):
\begin{equation}
\label{eq:H_PV_FW}
H_{\mathrm{FW}}^{\mathrm{PV}}
= \left(\frac{4\pi}{3}\right)^{1/3}\,\Gamma\!\left(\frac{5}{3}\right)\,\lambda^{1/3}\,v_c.
\end{equation}
\end{itemize}
\end{corollary}

Corollary~\ref{cor:HO_PV_3D} shows that the handover rate scales as $\lambda^{1/3}$, reflecting the fact that the density of Voronoi cell boundaries grows with the cube root of the BS density in 3D space. This highlights a key difference from 2D networks, where boundary density scales as $\lambda^{1/2}$.

\subsubsection{Specialization of the Sojourn-Time Law}

We next specialize the sojourn-time analysis. This requires the distribution of $D$ associated with Poisson--Voronoi cells.

\begin{corollary}[Approximate Mean Sojourn Time in 3D Poisson--Voronoi Networks]
\label{cor:pv_forward_sojourn}
Consider a stationary 3D Poisson--Voronoi tessellation generated by a homogeneous PPP $\Phi\subset\mathbb{R}^3$ with intensity $\lambda>0$. Let $D\ge 0$ denote the forward distance from a typical interior point to the first intersection with the Voronoi cell boundary along the current direction, and assume $\mathbb{E}[D]<\infty$. Under the cruise--turn mobility with cruise speed $v_c$, turn speed $v_t$, reorientation triggering rate $\nu$, and maximum angular speed $\omega_{\max}$, the mean sojourn time is approximated by \eqref{eq:meanTf_general}. The forward-distance PDF $f_D(d)$ for the 3D Poisson--Voronoi tessellation follows the classical result derived in \cite{Muche_2010} and is given by
\begin{align}
f_D(d)
=&\;
8\pi^2\lambda^2
\int_{0}^{\pi}\int_{0}^{\pi-\alpha}
\frac{d^5\sin^4\alpha\sin^2\beta}{\sin^6(\alpha+\beta)}
\left(b_0(\beta)+b_1(\beta)\right) \nonumber
\\
&\times\exp\!\left\{-\lambda v_3(d,\alpha,\beta)\right\}
\, \mathrm{d}\beta\, \mathrm{d}\alpha,
\label{eq:pv_forward_distance_pdf}
\end{align}
where
\begin{equation*}
b_0(\beta)=\cos^2\!\frac{\beta}{2},
\qquad
b_1(\beta)=-\frac{1}{4}\sin^2\beta,
\end{equation*}
and
\begin{equation*}
v_3(d,\alpha,\beta)
= \frac{4\pi}{3}\Big(\rho^3\big(a_0(\alpha)+a_1(\alpha)\big) + L^3\big(a_0(\beta)+a_1(\beta)\big)\Big),
\end{equation*}
with $a_0(\psi)=\cos^2\frac{\psi}{2}$, $a_1(\psi)=\frac14\sin^2\psi\cos\psi$, $\rho=\frac{d\sin\beta}{\sin(\alpha+\beta)}$, and $L=\sqrt{d^2+\rho^2-2d\rho\cos\alpha}$.
\end{corollary}

Unlike the handover rate, which depends only on $S_V$, the sojourn time involves the full distribution of the forward boundary distance. The integral expression reflects the intrinsic geometric complexity of Poisson--Voronoi cells, whose shape variability leads to a nontrivial distance distribution.

\subsection{Poisson--Delaunay Cellular Networks}
\label{subsec:pd_application}

\subsubsection{Geometry and Association Model}

We next consider Poisson--Delaunay tessellations, the dual of Voronoi tessellations. This model is particularly relevant to CoMP transmission, where multiple BSs jointly serve a user \cite{9151343,8976426,10720695}.

Each Delaunay cell is a tetrahedron whose vertices correspond to cooperating BSs:
\begin{equation}
\mathcal{S}(\mathbf x) = \mathrm{Vert}(C(\mathbf x)), \qquad |\mathcal{S}(\mathbf x)| = 4.
\end{equation}

Thus, boundary crossings correspond to changes in the cooperative BS set.

\begin{rmk}[Scope of the Poisson--Delaunay CoMP Architecture]
In the considered Poisson--Delaunay architecture, a user is served by the four vertices of its containing Delaunay tetrahedron; hence, \(|\mathcal S(\mathbf x)|=4\). Crossing a shared tetrahedral facet replaces one serving BS while retaining the other three. This construction therefore represents a specific fixed-cardinality, tessellation-based CoMP architecture. Other cooperative association rules, such as \(k\)-nearest-BS association and adaptive clustering, generally produce different service regions and serving-set-update boundaries. The general framework remains applicable when the resulting boundary process satisfies the stated conditions, with its specific surface area and forward boundary-distance distribution evaluated accordingly.
\end{rmk}

\subsubsection{Specialization of the Handover Law}

We now apply the universal handover law.

\begin{corollary}[Handover Rate in 3D Poisson--Delaunay Networks]
\label{cor:HO_PD_3D}
Consider a 3D Poisson--Delaunay tessellation generated by a homogeneous PPP of intensity $\lambda$. The specific surface area of the Delaunay cell boundaries is given in stochastic geometry \cite[Table 5.11.2]{okabe2000spatial} as
\begin{equation}
\label{eq:SV_PD}
S_V^{\mathrm{PD}}
= \frac{200}{27}\,\Gamma\!\left(\frac{2}{3}\right)\left(\frac{\pi\lambda}{6}\right)^{\!1/3}.
\end{equation}
Under the general cruise--turn mobility model, the handover rate is given by
\begin{equation}
\label{eq:H_PD_general}
H
= \frac{S_V^{\mathrm{PD}}}{2}\,\frac{\frac{v_c}{\nu} + v_t \dfrac{\pi}{2\omega_{\max}}}{\dfrac{1}{\nu} + \dfrac{\pi}{2\omega_{\max}}}.
\end{equation}

In particular,
\begin{itemize}
\item Multirotor UAVs (pause-to-turn, $v_t=0$):
\begin{equation}
\label{eq:H_PD_MR}
H_{\mathrm{MR}}^{\mathrm{PD}}
= \frac{100}{27}\,\Gamma\!\left(\frac{2}{3}\right)\left(\frac{\pi\lambda}{6}\right)^{\!1/3}\frac{v_c}{1+\nu\dfrac{\pi}{2\omega_{\max}}}.
\end{equation}

\item Fixed-wing UAVs (fly-to-turn, $v_t=v_c$):
\begin{equation}
\label{eq:H_PD_FW}
H_{\mathrm{FW}}^{\mathrm{PD}}
= \frac{100}{27}\,\Gamma\!\left(\frac{2}{3}\right)\left(\frac{\pi\lambda}{6}\right)^{\!1/3}\,v_c.
\end{equation}
\end{itemize}
\end{corollary}

Compared to Corollary~\ref{cor:HO_PV_3D}, although both Poisson--Voronoi and Poisson--Delaunay networks follow the same universal handover law, their geometric factors differ. This leads to different handover rates under identical mobility conditions, reflecting the impact of the association policy on boundary geometry.

\subsubsection{Specialization of the Sojourn-Time Law}

We now analyze the sojourn time. In contrast to the Voronoi case, an exact closed-form expression for the distribution of $D$ is not available. We therefore construct an approximation from its exact initial hazard rate.

\begin{lemma}[Linear Forward Boundary Distance Approximation]
\label{lem:pd_lcd}
Let $Z \subset \mathbb{R}^3$ denote the boundary random closed set induced by a 3D Poisson--Delaunay tessellation, and let $D$ be the forward boundary distance along an isotropically distributed direction. Define the linear contact distribution
\begin{equation}
F_D(r) \triangleq \mathbb{P}(D \le r), \qquad
f_D(r) \triangleq \frac{\mathrm d}{\mathrm dr} F_D(r),
\end{equation}
and the corresponding hazard rate as
\begin{equation}
\eta_D(r) \triangleq \frac{f_D(r)}{1 - F_D(r)}, \qquad r>0.
\end{equation}
Then the hazard rate satisfies
\begin{equation}
\eta_D(0+) = \frac{S_V^{\mathrm{PD}}}{2},
\end{equation}
where $S_V^{\mathrm{PD}}$ denotes the specific surface area of the boundary set $Z$. Moreover, under the constant-hazard approximation
\begin{equation}
\eta_D(r) \approx \eta_D(0+),
\end{equation}
the linear forward boundary distance admits the approximate exponential density
\begin{equation}
\label{eq:lcd_pdf}
f_D(r)
\approx \frac{S_V^{\mathrm{PD}}}{2}\,\exp\!\left(-\frac{S_V^{\mathrm{PD}}}{2}\, r\right),
\qquad r\ge 0.
\end{equation}
\end{lemma}

\begin{IEEEproof}
For stationary face-to-face tessellations (including Poisson--Delaunay tessellations), the linear contact distribution satisfies $F_D(r)=H_{\ell}(r)$. Integral-geometric arguments yield $H_{\ell}'(0+)=P_L$, where $P_L$ denotes the mean number of intersections of a unit-length isotropic line segment with $Z$. By the stereological Crofton identity for stationary isotropic boundary sets, $S_V^{\mathrm{PD}} = 2P_L$, which implies $\eta_D(0+) = S_V^{\mathrm{PD}}/2$.

The constant-hazard approximation replaces $\eta_D(r)$ by its exact initial value, i.e., $\eta_D(r)\approx \eta_D(0+)$. Substituting this approximation into the definition of the hazard rate gives
\begin{equation}
\frac{\mathrm d}{\mathrm dr} F_D(r)
= \frac{S_V^{\mathrm{PD}}}{2}\,[1 - F_D(r)],
\end{equation}
with the initial condition $F_D(0)=0$. Solving this differential equation yields $F_D(r)=1-\exp(-\tfrac{S_V^{\mathrm{PD}}}{2}r)$, and differentiation gives the stated density.
\end{IEEEproof}

\begin{figure}[!t]
  \centering
  \includegraphics[width=3in]{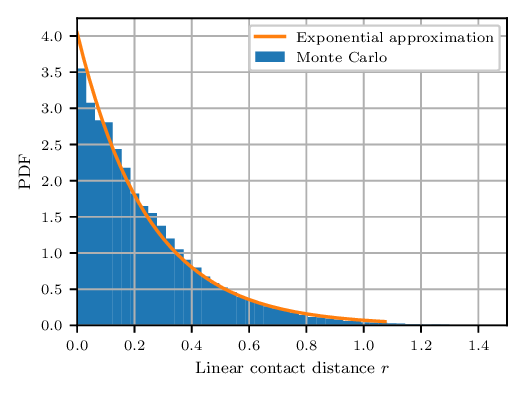}
  \vspace{-10pt}
\caption{PDF of the linear forward boundary distance for a 3D Poisson--Delaunay tessellation, obtained via Monte Carlo simulation and compared with the exponential approximation.}
  \label{fig:pd_linear_contact_pdf}
\end{figure}

\begin{rmk}[Exponential Approximation]
The constant-hazard approximation implies that the forward boundary distance is approximately exponentially distributed. Because it is anchored by the exact relation $\eta_D(0+)=S_V^{\mathrm{PD}}/2$, it reproduces the first-order behavior of the contact distribution at the origin. Figure~\ref{fig:pd_linear_contact_pdf} compares \eqref{eq:lcd_pdf} with Monte Carlo estimates for the 3D Poisson--Delaunay tessellation and assesses its accuracy beyond this local regime. The close agreement over the displayed range provides numerical support for using the approximation in the mean-sojourn-time calculation.
\end{rmk}

Using the exponential approximation for the forward boundary-distance distribution in Lemma~\ref{lem:pd_lcd}, we now specialize the general sojourn-time expression \eqref{eq:meanTf_general} to the Poisson--Delaunay tessellation.

\begin{corollary}[Approximate Mean Sojourn Time in 3D Poisson--Delaunay Networks]
\label{cor:pd_forward_sojourn_from_lemma}
Under the setting of Lemma~\ref{lem:pd_lcd}, consider the cruise--turn mobility model with cruise speed $v_c$, turn speed $v_t$, reorientation triggering rate $\nu$, and maximum angular speed $\omega_{\max}$. Let $T_f$ denote the remaining sojourn time, i.e., the time required to reach the first intersection with the Poisson--Delaunay cell boundary along the current direction. Then, under the constant-hazard approximation in \eqref{eq:lcd_pdf}, the forward boundary distance \(D\) is approximated by an exponential distribution with \(\mathbb E[D]\approx 2/S_V^{\mathrm{PD}}\). Combining this with the effective-traversal approximation of Theorem~\ref{thm:SojournTime} gives
\begin{equation}
\mathbb{E}[T_f] \approx \left(\frac{1}{v_c} + \frac{\pi \kappa (1-\zeta)}{2\omega_{\max}}\right)\frac{2}{S_V^{\mathrm{PD}}}.
\label{eq:pd_Tf_mean_closed}
\end{equation}
\end{corollary}

Within the adopted approximations, \eqref{eq:pd_Tf_mean_closed} reveals a reciprocal dependence on \(S_V^{\mathrm{PD}}\). Thus, to first order, the same specific surface area that governs the handover rate also describes the local boundary geometry relevant to sojourn time: increasing boundary density raises the handover frequency while shortening the mean remaining sojourn time.

\subsection{Summary of Applications}

The two case studies illustrate how the unified framework specializes to concrete tessellation models, translating geometric properties into explicit mobility performance metrics. In particular, the exact handover rate depends solely on the specific surface area, which serves as a global geometric descriptor, whereas the mean-sojourn-time approximation uses the forward boundary distance distribution to capture local geometric variability. These results demonstrate how the same framework can be applied across different 3D cellular architectures while retaining a clear distinction between exact and approximate results.

\section{Numerical Results and Discussion} \label{sec:numerical_results}

In this section, Monte Carlo simulations corroborate the exact handover-rate expressions, assess the accuracy of the mean-sojourn-time approximations, and quantify the effects of deployment geometry and UAV mobility. We consider three representative service tessellations: rhombic dodecahedral \cite{Alam2006Coverage}, Poisson--Voronoi, and Poisson--Delaunay. The final subsection illustrates how the derived handover laws can guide BS-density and service-architecture selection.

Unless otherwise specified, the parameters are listed in Table~\ref{tab:simulation_parameters}. Two mobility regimes are considered: multirotor UAVs with pause-to-turn motion (\(v_t=0\)) and fixed-wing UAVs with fly-to-turn motion (\(v_t=v_c\)). Each Monte Carlo estimate is based on \(N_{\mathrm{MC}}=200\) independent realizations.

\begin{table}[!t]
\centering
\renewcommand{\arraystretch}{1.08}
\caption{Default simulation and link-level parameters.}
\label{tab:simulation_parameters}
{\footnotesize
\setlength{\tabcolsep}{6pt}
\begin{tabular}{!{\vrule width1.2pt}>{\centering\arraybackslash}m{0.42\columnwidth}|>{\centering\arraybackslash}m{0.46\columnwidth}!{\vrule width1.2pt}}
\noalign{\hrule height 1.2pt}
\textbf{Parameter} & \textbf{Value} \\
\noalign{\hrule height 1.2pt}

Cellular deployment & Rhombic dodecahedron, Poisson--Voronoi, and Poisson--Delaunay \\
\hline
Reference BS density, $\lambda$ & $3\times10^{2}$~\si{BS/km^3} \\
\hline
Cruise speed, $v_c$ & $25$~\si{m/s} \\
\hline
Reorientation triggering rate, $\nu$ & $0.18$~\si{s^{-1}} \\
\hline
Maximum angular speed, $\omega_{\max}$ & $0.35$~\si{rad/s} \\
\hline
Turn speed, $v_t$ & $0$ (multirotor) or $v_c$ (fixed-wing) \\
\hline
Carrier frequency, $f_c$ & $3.5$~\si{GHz} \\
\hline
Transmit power, $P_t$ & $21$~\si{dBm} \\
\hline
Bandwidth, $B$ & $50$~\si{MHz} \\
\hline
Noise figure & $7$~\si{dB} \\
\hline
Path-loss exponent, $\alpha$ & $3$ \\
\hline
Mean service-interruption duration, $\tau_a$ & $0.5$~\si{s} \\
\hline
Target SINR threshold, $\theta$ & $0$~\si{dB} \\
\hline
Density-search range, $\Lambda$ & $[30,3\times10^{4}]$~\si{BS/km^3} \\
\hline
Monte Carlo realizations, $N_{\mathrm{MC}}$ & $200$ \\
\noalign{\hrule height 1.2pt}
\end{tabular}
}
\end{table}

\subsection{Handover Rate in Different Cellular Networks}

For the Poisson--Voronoi and rhombic-dodecahedral deployments, a handover occurs when the UAV crosses a cell boundary and changes its serving BS. For the Poisson--Delaunay architecture, a boundary crossing updates the four-BS cooperative serving set: crossing a shared tetrahedral facet replaces one BS while retaining the other three.

\begin{figure*}[!t]
  \centering
  \captionsetup[subfigure]{font=scriptsize,labelfont={scriptsize,sf},textfont={scriptsize,sf},skip=2pt}
  \subfloat[Handover rate $H$ versus BS density $\lambda$.\label{fig:handover_lambda}]{%
    \includegraphics[width=0.32\textwidth]{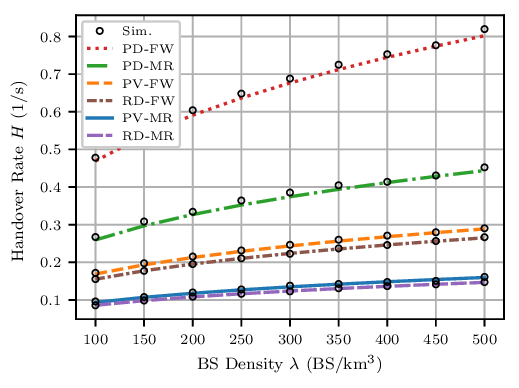}}
\hfill
  \subfloat[Handover rate $H$ versus reorientation triggering rate $\nu$ for multirotor UAVs.\label{fig:handover_nu}]{%
    \includegraphics[width=0.32\textwidth]{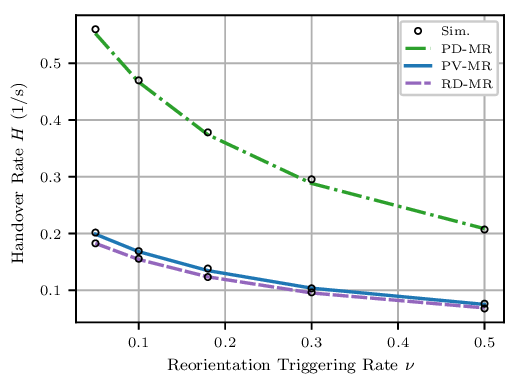}}
\hfill
  \subfloat[Handover rate $H$ versus UAV speed $v_c$.\label{fig:handover_vc}]{%
    \includegraphics[width=0.32\textwidth]{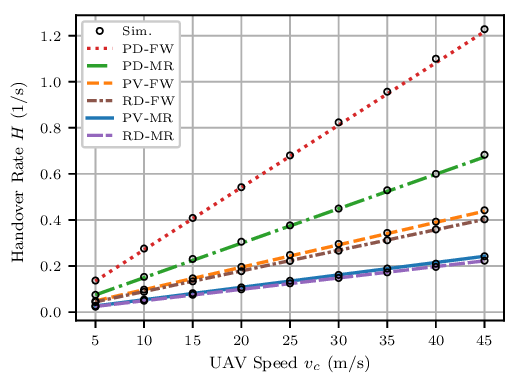}}
\caption{Handover rate under different deployments and mobility regimes. The rate increases with BS density and cruise speed, and decreases with the multirotor reorientation triggering rate because $H\propto\bar v$.}
  \label{fig:handover_combined}
\end{figure*}

Fig.~\ref{fig:handover_combined}\subref{fig:handover_lambda} shows that \(H\) increases with \(\lambda\) for every deployment and mobility regime. For fixed \(\bar v\), this trend follows from \(H\propto S_V\propto\lambda^{1/3}\). The Poisson--Delaunay architecture has the largest specific surface area and hence the highest handover rate, followed by Poisson--Voronoi and rhombic-dodecahedral service. Fixed-wing UAVs experience more frequent handovers because their fly-to-turn motion produces a larger long-run mean speed. The close agreement between the analytical curves and Monte Carlo results corroborates the exact handover expressions.

For multirotor UAVs, Fig.~\ref{fig:handover_combined}\subref{fig:handover_nu} shows that increasing \(\nu\) reduces \(H\). More frequent zero-translation turns increase the fraction of time spent reorienting and thereby reduce \(\bar v\). In contrast, Fig.~\ref{fig:handover_combined}\subref{fig:handover_vc} confirms that \(H\) increases linearly with \(v_c\). Across both parameter variations, the deployment ordering remains unchanged because mobility determines \(\bar v\), whereas the service geometry determines \(S_V\).

At a matched long-run mean speed, an admissible continuous random-direction or random-walk baseline yields the same exact handover rate and would therefore coincide with the corresponding curves shown here. The distinction between the multirotor and fixed-wing results arises from how their platform parameters map to \(\bar v\), rather than from their directional statistics alone.

\subsection{Sojourn Time in Different Cellular Networks}

\begin{figure*}[!t]
  \centering
  \captionsetup[subfigure]{font=scriptsize,labelfont={scriptsize,sf},textfont={scriptsize,sf},skip=2pt}
  \subfloat[Mean sojourn time ${\mathbb{E}[T_f]}$ versus BS density $\lambda$.\label{fig:sojourn_lambda}]{%
    \includegraphics[width=0.32\textwidth]{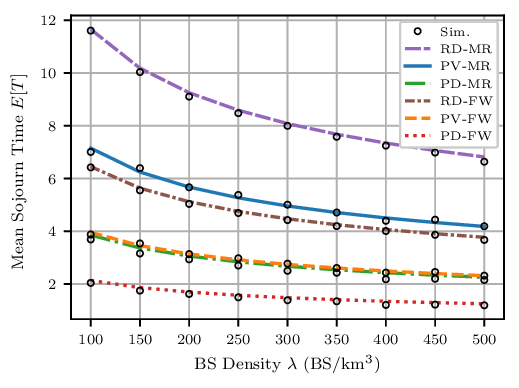}}
\hfill
  \subfloat[Mean sojourn time ${\mathbb{E}[T_f]}$ versus reorientation triggering rate $\nu$ for multirotor UAVs.\label{fig:sojourn_nu}]{%
    \includegraphics[width=0.32\textwidth]{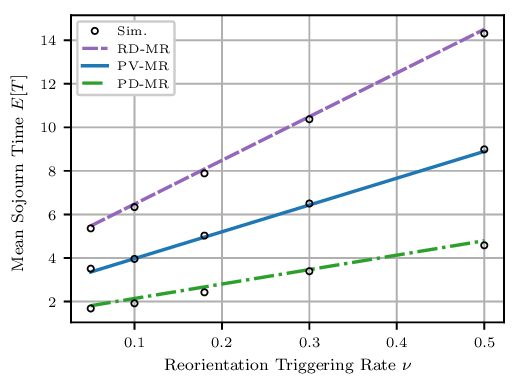}}
\hfill
  \subfloat[Mean sojourn time ${\mathbb{E}[T_f]}$ versus UAV speed $v_c$.\label{fig:sojourn_vc}]{%
    \includegraphics[width=0.32\textwidth]{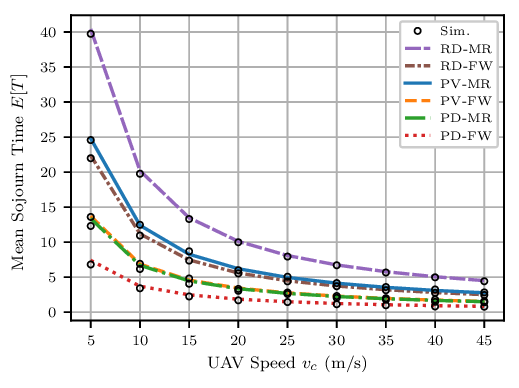}}
\caption{Approximate mean sojourn time under different deployments and mobility regimes. Markers denote Monte Carlo results, and curves denote the analytical approximations.}
  \label{fig:sojourn_combined}
\end{figure*}

Fig.~\ref{fig:sojourn_combined} compares Monte Carlo estimates with the mean-sojourn-time approximations. As shown in Fig.~\ref{fig:sojourn_combined}\subref{fig:sojourn_lambda}, \(\mathbb E[T_f]\) decreases with \(\lambda\), consistently with the increasing density of service boundaries. Poisson--Delaunay service gives the shortest mean sojourn time, whereas the rhombic-dodecahedral deployment gives the longest. Fixed-wing UAVs also have shorter sojourn times because their fly-to-turn motion maintains translation during reorientation.

For multirotor UAVs, Fig.~\ref{fig:sojourn_combined}\subref{fig:sojourn_nu} shows that \(\mathbb E[T_f]\) increases with \(\nu\), because more frequent pause-to-turn events increase the effective traversal time. Conversely, Fig.~\ref{fig:sojourn_combined}\subref{fig:sojourn_vc} shows that increasing \(v_c\) shortens the time required to reach the next boundary. The agreement with the Monte Carlo results supports the effective-traversal approximation over the parameter ranges considered.

\subsection{Handover-Aware Effective Spectral Efficiency and BS-Density Optimization}
\label{subsec:density_optimization}

We finally evaluate the handover-aware effective spectral efficiency in \eqref{eq:Ga_general} and the BS-density selection rule in \eqref{eq:lambda_star_general}. Under Poisson--Voronoi service, an aerial user associates with its nearest BS; the coverage probability is taken from \cite{7058418}, and the handover rate is given by Corollary~\ref{cor:HO_PV_3D}. Under Poisson--Delaunay service, the four vertices of the containing tetrahedron form the cooperative serving set according to the noncoherent CoMP model in \cite{9151343}, and the serving-set-update rate is given by Corollary~\ref{cor:HO_PD_3D}.

The reference comparison assumes equal effective antenna gains for the desired and interfering links. Directional or elevation-dependent gains would modify link coverage and, under gain-based association, could also alter the service-region geometry. Such antenna models can be incorporated by recomputing the corresponding coverage and boundary descriptors. The following results therefore pertain to the equal-gain reference model.

\begin{figure}[!t]
\centering
\captionsetup[subfigure]{font=scriptsize,labelfont={scriptsize,sf},textfont={scriptsize,sf},skip=2pt}
\subfloat[Poisson--Voronoi single-BS association.\label{fig:Ga_PV}]{%
\includegraphics[width=3in]{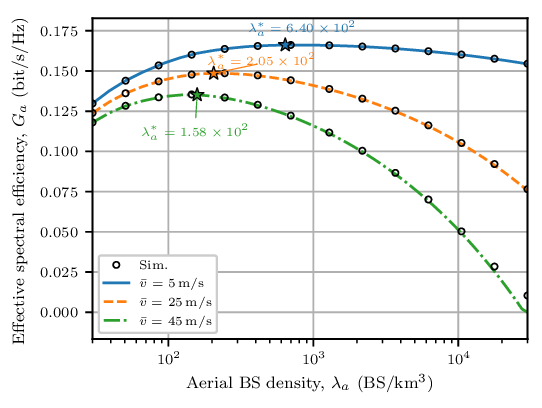}}\\[1mm]
\subfloat[Poisson--Delaunay four-BS noncoherent CoMP.\label{fig:Ga_PD}]{%
\includegraphics[width=3in]{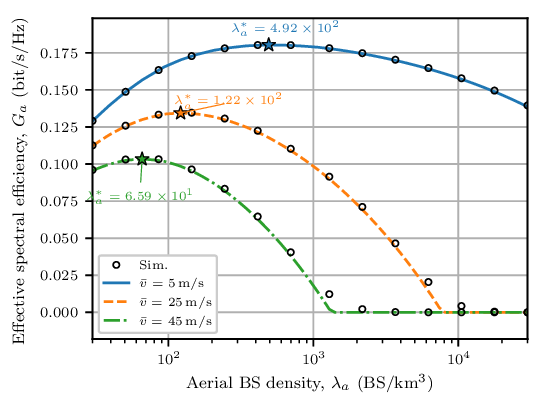}}
\caption{Handover-aware effective spectral efficiency \(G_a\) versus BS density for \(\theta=0\)~\si{dB} and \(\tau_a=0.5\)~\si{s}. Open markers denote Monte Carlo results, and stars mark the density optima obtained by grid search.}
\label{fig:Ga_density}
\end{figure}

Fig.~\ref{fig:Ga_density} shows close agreement between the analytical predictions, obtained by combining the adopted coverage models with the exact handover laws, and the Monte Carlo results. Define the normalized handover coefficient as
\[
c_{\mathcal T}
=
\frac{H^{\mathcal T}(\lambda,\bar v)}
{\lambda^{1/3}\bar v}.
\]
The simulated values are \(1.447\pm0.031\) and \(3.997\pm0.066\) for Poisson--Voronoi and Poisson--Delaunay service, respectively, compared with the analytical values \(1.455\) and \(4.042\). The reported uncertainties are the half-widths of the corresponding \(95\%\) confidence intervals. For each speed, \(G_a\) initially increases with BS density because of improved coverage, but eventually decreases as the coverage gain saturates while the interruption penalty continues to grow with \(\lambda^{1/3}\).

For Poisson--Voronoi service, the optimizing densities are approximately \(6.40\times10^2\), \(2.05\times10^2\), and \(1.58\times10^2\)~\si{BS/km^{3}} for \(\bar v=5\), \(25\), and \(45\)~\si{m/s}, respectively. The corresponding Poisson--Delaunay values are \(4.92\times10^2\), \(1.22\times10^2\), and \(6.59\times10^1\)~\si{BS/km^{3}}. The respective maxima of \(G_a\) are \(0.166\), \(0.149\), and \(0.135\)~\si{bit/s/Hz} for Poisson--Voronoi service and \(0.180\), \(0.134\), and \(0.103\)~\si{bit/s/Hz} for Poisson--Delaunay service. Thus, four-BS CoMP provides the higher \(G_a\) at \(5\)~\si{m/s}, but the ordering reverses by \(25\)~\si{m/s}, and the single-BS advantage widens at \(45\)~\si{m/s}. In both architectures, the optimizing density decreases with speed as the increasing handover cost makes dense deployment less favorable. This decrease is more pronounced for Poisson--Delaunay service, consistent with its larger normalized serving-set-update coefficient.

The preceding comparison assumes \(\tau_a=0.5\)~\si{s}. Because the interruption duration depends on the handover procedure and service architecture, we next examine the sensitivity of the design conclusions to \(\tau_a\). For each interruption duration, the BS density is reoptimized according to \eqref{eq:lambda_star_general}, so the comparison uses the best density available to each architecture.

\begin{figure}[!t]
\centering
\includegraphics[width=3in]{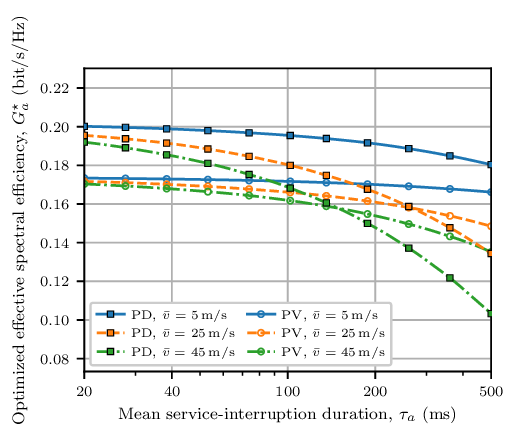}
\vspace{-10pt}
\caption{BS-density-optimized handover-aware effective spectral efficiency versus mean service-interruption duration \(\tau_a\). PD and PV denote Poisson--Delaunay four-BS noncoherent CoMP and Poisson--Voronoi single-BS association, respectively.}
\label{fig:Ga_tau_sensitivity}
\end{figure}

Fig.~\ref{fig:Ga_tau_sensitivity} shows that the density-optimized effective spectral efficiency decreases monotonically with \(\tau_a\), despite reoptimization of the BS density. At \(\bar v=5\)~\si{m/s}, Poisson--Delaunay service remains superior throughout the considered range because its CoMP coverage gain outweighs the serving-set-update cost. At \(\bar v=25\)~\si{m/s}, the two architectures become comparable at approximately \(\tau_a=270\)~\si{ms}, beyond which Poisson--Voronoi service performs better. At \(\bar v=45\)~\si{m/s}, the crossover occurs earlier, at approximately \(\tau_a=150\)~\si{ms}. This earlier crossover follows from the linear growth of the update rate with user speed and the larger normalized update coefficient of the Poisson--Delaunay architecture. Hence, CoMP is favored by low mobility and short interruptions, whereas single-BS service becomes increasingly advantageous as either quantity grows.

\begin{figure}[!t]
\centering
\includegraphics[width=3in]{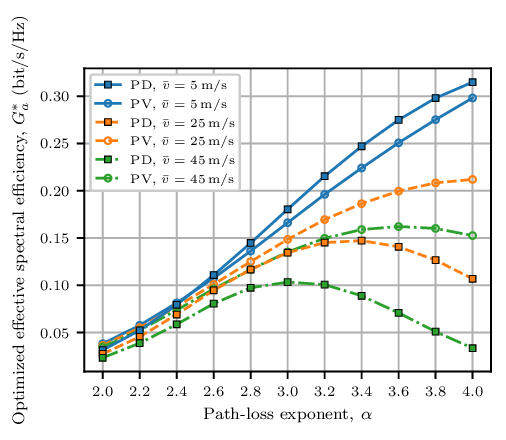}
\caption{BS-density-optimized effective spectral efficiency versus path-loss exponent \(\alpha\). The BS density is reoptimized for each value of \(\alpha\).}
\label{fig:alpha_sensitivity}
\end{figure}

Propagation conditions also affect the relative benefit of cooperation. To examine this effect, Fig.~\ref{fig:alpha_sensitivity} varies the path-loss exponent \(\alpha\), with the BS density reoptimized for each value of \(\alpha\) and the remaining parameters set according to Table~\ref{tab:simulation_parameters}.

Fig.~\ref{fig:alpha_sensitivity} shows that the effect of \(\alpha\) depends on both user speed and service architecture. At low mobility, both architectures benefit from a larger \(\alpha\), and Poisson--Delaunay service becomes superior at moderate and large values. At higher speeds, however, its performance first increases and then decreases because stronger attenuation weakens the non-nearest cooperative links and reduces the CoMP gain, while the serving-set-update cost remains comparatively high. Poisson--Voronoi service is less sensitive to this loss because it relies only on the nearest serving BS.

Collectively, these results show that BS density and service architecture are coupled design variables. Poisson--Delaunay CoMP is favored when mobility and interruption overhead are sufficiently low for its cooperative coverage gain to outweigh the higher serving-set-update cost. Poisson--Voronoi service becomes preferable as user speed, interruption duration, or attenuation of the non-nearest cooperative links increases. The BS density and service architecture should therefore be optimized jointly rather than selected independently.

\section{Conclusion}
\label{sec:conclusion}

This paper developed a unified geometric--kinematic framework for analyzing mobility-induced handovers and sojourn behavior in 3D cellular networks. Under the stated stationarity, isotropy, and independence conditions, the handover rate obeys an exact law determined by the specific surface area of the service boundary and the long-run mean user speed, whereas the approximate mean remaining sojourn time depends on the forward boundary-distance distribution through an effective-traversal model. Specialization to Poisson--Voronoi and Poisson--Delaunay networks yielded exact handover-rate and approximate mean-sojourn-time expressions for single-BS and four-BS CoMP service, respectively. Incorporating the handover law into a handover-aware effective spectral efficiency further showed that BS density and service architecture are coupled design variables whose joint selection depends on mobility, interruption duration, and propagation conditions. Potential extensions include repulsive, inhomogeneous, and anisotropic deployments; directional antennas and gain-based association; and joint temporal modeling of boundary crossings, service interruptions, and SINR. More detailed trajectory-dependent modeling of the boundary first-exit process may also improve the sojourn-time approximation.

\appendices

\section{Proof of Lemma \ref{lem:degenerate_hits}}\label{sec:degenerate_hits_proof}

Since $\mathcal{C}$ is a face-to-face polyhedral tessellation, its boundary $Z$ can be decomposed as
\begin{equation}
Z = Z^{(2)} \,\dot\cup\, Z^{(1)} \,\dot\cup\, Z^{(0)},
\end{equation}
where $Z^{(2)}$ is the union of the relative interiors of all facets, $Z^{(1)}$ is the union of all edges, and $Z^{(0)}$ is the set of all vertices. In any bounded observation window, $Z^{(1)}$ consists of finitely many line segments and $Z^{(0)}$ consists of finitely many points.

Fix a finite time horizon $[0, T]$ and condition on a realization of the tessellation $\mathcal C$. Since the trajectory $\{\mathbf X(t)\}_{0\le t\le T}$ is absolutely continuous, its image $\Gamma_T \triangleq \{\mathbf X(t):0\le t\le T\}$ is a rectifiable curve of finite length.

We first consider intersections with $Z^{(1)}$. For a fixed realization of the motion after time $0$, the event $\Gamma_T \cap Z^{(1)} \neq \emptyset$ implies that the initial point $\mathbf X(0)$ belongs to the translate of $Z^{(1)}$ by some point of the relative displacement set generated by the trajectory over $[0, T]$. Equivalently, the set of initial locations leading to an intersection with $Z^{(1)}$ is contained in $Z^{(1)} \oplus (-\Gamma_T^{\rm rel})$, where $\Gamma_T^{\rm rel}$ is the path traced relative to the initial point. Since $Z^{(1)}$ is one-dimensional and $\Gamma_T^{\rm rel}$ is rectifiable and one-dimensional, their Minkowski sum is contained in a countable union of two-dimensional surfaces and therefore has zero 3D Lebesgue measure. Because the law of $\mathbf X(0)$ is absolutely continuous with respect to Lebesgue measure, it follows that $\mathbb P\!\left(\Gamma_T \cap Z^{(1)} \neq \emptyset \,\middle|\, \mathcal C,\Gamma_T^{\rm rel}\right)=0$. Hence, $\mathbb P\!\left(\Gamma_T \cap Z^{(1)} \neq \emptyset\right)=0$.

The same argument applies to $Z^{(0)}$, since $Z^{(0)}$ is zero-dimensional, and thus $\mathbb P\!\left(\Gamma_T \cap Z^{(0)} \neq \emptyset\right)=0$. Therefore, $\mathbb P\!\left(\Gamma_T \cap (Z^{(1)}\cup Z^{(0)}) \neq \emptyset\right)=0$ for every finite $T>0$.

Now let $T_n=n$ for $n\in\mathbb N$. Since
\begin{align}
\lefteqn{\left\{\exists\, t\ge 0:\mathbf X(t)\in Z^{(1)}\cup Z^{(0)}\right\}} \nonumber  \\
& =
\bigcup_{n=1}^\infty
\left\{\Gamma_{T_n}\cap (Z^{(1)}\cup Z^{(0)})\neq\emptyset\right\}, \nonumber
\end{align}
a countable union bound yields
\begin{equation}
\mathbb P\Big(\exists\, t\ge 0:\mathbf X(t)\in Z^{(1)}\cup Z^{(0)}\Big)=0.
\end{equation}
Therefore, almost surely, every boundary hit satisfies $\mathbf X(t)\in Z^{(2)}$, i.e., every intersection occurs in the relative interior of a facet.

It remains to exclude tangential contacts with facet interiors. Conditional on a facet with unit normal $\mathbf n$, tangency during a cruise segment requires the flight direction to lie in the great circle \(\{\mathbf u\in\mathbb S^2:\langle\mathbf u,\mathbf n\rangle=0\}\), which has zero spherical measure. During a nondegenerate turn, the position and velocity depend smoothly on the continuously distributed endpoint direction; simultaneous contact with the supporting plane and zero normal velocity likewise defines a parameter set of spherical measure zero. In the pause-to-turn case, a stationary turn could occur on a facet only if the preceding cruise ended exactly on that facet, an event of probability zero. There are only countably many motion phases and, locally, countably many facets. Independence and the directional-density assumption therefore imply that tangential facet contacts occur with probability zero.

Thus every boundary contact is almost surely a transverse crossing through the relative interior of a single facet. In a face-to-face tessellation, such a crossing leaves exactly one cell and enters exactly one adjacent cell. Consequently, each transverse intersection produces one service-domain change, and the handover count equals the number of transverse intersections with $Z$, counted with multiplicity along the trajectory. This completes the proof.

\section{Proof of Theorem \ref{thm:handover}} \label{sec:proof_handover}

By Lemma~\ref{lem:degenerate_hits}, degenerate and tangential contacts occur with probability zero, and each transverse boundary intersection produces exactly one service-domain change. Hence,
\begin{equation}
N_{\mathrm{HO}}(T)=N_Z\!\big(\mathbf X;[0,T]\big)\qquad \text{a.s.},
\label{eq:HO_equals_intersections_again}
\end{equation}
where $N_Z(\mathbf X;[0,T])$ denotes the number of transverse intersections with $Z$, counted with multiplicity along the trajectory.

Let
\begin{equation}
L(T)\triangleq \int_0^T V(t)\,\mathrm{d}t
\label{eq:path_length_over_T}
\end{equation}
denote the traveled arc length over $[0,T]$.

Condition on the realized trajectory over $[0,T]$. Because the trajectory and $Z$ are independent, the conditional path is a deterministic rectifiable curve. For a stationary isotropic random surface, the mean number of transverse intersections with a deterministic rectifiable curve equals its arc length multiplied by the line-intersection intensity $P_L$ \cite[Ch.~9]{schneider2008stochastic}. This identity follows first for polygonal curves by additivity over their segments and then for rectifiable curves by polygonal approximation; counting with multiplicity also covers self-intersecting paths. Therefore,
\begin{equation}
\mathbb{E}\!\left[N_Z\!\big(\mathbf X;[0,T]\big)\,\middle|\,
\{\mathbf X(t):0\le t\le T\}\right]
=P_L L(T).
\label{eq:EN_given_path}
\end{equation}

Taking expectations and using \eqref{eq:HO_equals_intersections_again} yields
\begin{equation*}
\mathbb{E}[N_{\mathrm{HO}}(T)]
=P_L\,\mathbb{E}[L(T)]
=P_L\,\mathbb{E}\!\left[\int_0^T V(t)\,\mathrm{d}t\right].
\end{equation*}

By stationarity of the velocity process and Fubini's theorem,
\begin{equation}
\frac{1}{T}\mathbb{E}\!\left[\int_0^T V(t)\,\mathrm{d}t\right]
=\mathbb E[V(0)].
\label{eq:ergodic_speed_identity}
\end{equation}
The ergodic theorem identifies \(\mathbb E[V(0)]=\bar v\) almost surely. Dividing by \(T\) and passing to the limit thus gives \(H=P_L\bar v\).

For a stationary isotropic boundary set $Z$, contact distribution theory gives $H_\ell'(0^+)=P_L$ \cite[Sec.~11.2]{schneider2008stochastic}, while the stereological Crofton identity gives $S_V = 2P_L$ \cite[Ch.~9]{schneider2008stochastic} \cite[Eq.~(24)]{105566}. Hence,
\begin{equation}
\label{eq:crofton_identity}
H_\ell'(0^+) = P_L = \frac{S_V}{2},
\end{equation}
and substitution into \(H=P_L\bar v\) yields \eqref{eq:handover_rate_main}.

\section{Proof of Lemma \ref{lem:vbar}}\label{sec:vbar_proof}

For the $k$-th cruise--turn cycle, define
\begin{equation*}
C_k \triangleq T_{c,k}+\tau_k,
\qquad
R_k \triangleq v_c T_{c,k}+v_t\tau_k,
\end{equation*}
where $C_k$ denotes the cycle duration and $R_k$ denotes the traveled arc length during that cycle.

We first show that $\{(C_k, R_k)\}_{k\ge 1}$ are i.i.d. By assumption, $\{T_{c,k}\}_{k\ge1}$ are i.i.d.\ exponential random variables and are independent of the heading process. Moreover, the initial heading satisfies $\mathbf U_1\sim \mathrm{Unif}(\mathbb{S}^2)$, and for each $k\ge1$, the new heading $\mathbf U_{k+1}$ is sampled independently and uniformly on $\mathbb{S}^2$. Hence, $\{\mathbf U_k\}_{k\ge1}$ are i.i.d.\ isotropic directions.

For each $k\ge1$, define
\begin{equation*}
\Theta_k=\arccos\!\big(\langle \mathbf U_k,\mathbf U_{k+1}\rangle\big).
\end{equation*}
By rotational invariance, conditioning on any fixed $\mathbf U_{k+1}=\mathbf u\in\mathbb S^2$, the conditional distribution of $\Theta_k$ depends only on the isotropic distribution of $\mathbf U_{k}$ and is therefore independent of $\mathbf u$. Likewise, conditioned on $\mathbf U_{k+1}$, the variables $\Theta_k$ and $\Theta_{k+1}$ are independent, since they depend on $\mathbf U_{k}$ and $\mathbf U_{k+2}$, respectively, which are independent. Repeating this argument shows that $\{\Theta_k\}_{k\ge1}$ are i.i.d. Consequently,
\begin{equation*}
\tau_k=\frac{\Theta_k}{\omega_{\max}}
\end{equation*}
is also an i.i.d.\ sequence, and hence so is $\{(C_k,R_k)\}_{k\ge1}$, with finite means.

Now let
\begin{equation*}
L(T)\triangleq \int_0^T V(t)\,\mathrm{d}t
\end{equation*}
denote the traveled arc length over $[0,T]$, and let $N(T)$ be the number of completed cruise--turn cycles by time $T$. Then
\begin{equation*}
L(T)=\sum_{k=1}^{N(T)}R_k+\Delta(T),
\end{equation*}
where $\Delta(T)$ is the residual arc length accumulated during the possibly incomplete final cycle. Since at most one incomplete cycle contributes to $\Delta(T)$, the residual term is bounded by the arc length traveled during one cycle. Therefore,
\begin{equation*}
\frac{\Delta(T)}{T}\to 0
\qquad \text{a.s.}
\end{equation*}

By the renewal--reward theorem (see, e.g., Corollary~4.4.1 in \cite[Sec.~4.4.1, pp.~171--174]{Gallager2013}), we obtain
\begin{equation*}
\lim_{T\to\infty}\frac{L(T)}{T}
= \frac{\mathbb E[R]}{\mathbb E[C]}
\qquad \text{a.s.}
\end{equation*}
Since
\begin{equation*}
\bar v
= \lim_{T\to\infty}\frac{1}{T}\int_0^T V(t)\,\mathrm{d}t
= \lim_{T\to\infty}\frac{L(T)}{T},
\end{equation*}
it follows that
\begin{equation*}
\bar v
= \frac{\mathbb{E}[v_cT_c]+\mathbb{E}[v_t\tau]}
     {\mathbb{E}[T_c]+\mathbb{E}[\tau]}
\qquad \text{a.s.}
\end{equation*}

Since $T_c\sim \mathrm{Exp}(\nu)$,
\begin{equation*}
\mathbb{E}[T_c]=\frac{1}{\nu},
\qquad
\mathbb{E}[v_cT_c]=v_c\mathbb{E}[T_c]=\frac{v_c}{\nu},
\end{equation*}
and therefore
\begin{equation*}
\bar v
= \frac{\frac{v_c}{\nu}+v_t\,\mathbb{E}[\tau]}
     {\frac{1}{\nu}+\mathbb{E}[\tau]}.
\end{equation*}

It remains to compute $\mathbb{E}[\tau]$. Let $\mathbf U^{-},\mathbf U^{+}$ be two independent isotropic directions on $\mathbb{S}^2$, and define
\begin{equation*}
\Theta=\arccos\!\big(\langle \mathbf U^{-},\mathbf U^{+}\rangle\big)\in[0,\pi].
\end{equation*}
By rotational invariance, $\cos\Theta$ is uniformly distributed on $[-1,1]$. Equivalently, $\Theta$ has density
\begin{equation*}
f_\Theta(\theta)=\frac{1}{2}\sin\theta,\qquad \theta\in[0,\pi],
\end{equation*}
see, e.g., \cite{WeissteinSpherePointPicking}. Hence,
\begin{equation*}
\mathbb{E}[\Theta]
= \int_0^\pi \theta\,\frac{1}{2}\sin\theta\,\mathrm{d}\theta
= \frac{\pi}{2},
\end{equation*}
and therefore
\begin{equation*}
\mathbb{E}[\tau]
= \frac{\mathbb{E}[\Theta]}{\omega_{\max}}
= \frac{\pi}{2\omega_{\max}}.
\end{equation*}

This completes the proof.

\section{Proof of Theorem~\ref{thm:SojournTime}} \label{sec:proof_SojournTime}

Because the cruise durations are i.i.d.\ exponential random variables with rate $\nu$, reorientation triggers form a Poisson process of rate $\nu$ on the accumulated cruise-time axis. Accumulating cruise distance $d$ requires $d/v_c$ units of cruise time. Hence, the number of reorientation events that occur before accumulating cruise distance $d$ satisfies
\begin{equation}
N \mid (D=d) \sim \mathrm{Poisson}\!\left(\nu \frac{d}{v_c}\right) = \mathrm{Poisson}(\kappa d),
\end{equation}
where $\kappa=\nu/v_c$.

Each turn incurs an effective time overhead equal to the portion of the turn duration that does not contribute to translational progress, namely $(1-\zeta)\tau_i$, where $\zeta=v_t/v_c$. Therefore,
\begin{equation}
\mathbb E[T_f \mid D=d] = \frac{d}{v_c} + (1-\zeta)\, \mathbb E\!\left[\sum_{i=1}^{N}\tau_i \,\middle|\, D=d\right].
\end{equation}
Because $\{\tau_i\}$ are i.i.d.\ and independent of $N$, and $\mathbb E[N\mid D=d]=\kappa d$, we have
\begin{equation}
\mathbb E[T_f \mid D=d] = d\left(\frac{1}{v_c} + \kappa (1-\zeta)\,\mathbb E[\tau]\right).
\end{equation}

Using Lemma~\ref{lem:vbar}, which gives $\mathbb E[\tau]=\pi/(2\omega_{\max})$, we obtain
\begin{equation}
\mathbb E[T_f \mid D=d] = d \left(\frac{1}{v_c} + \frac{\pi\kappa(1-\zeta)}{2\omega_{\max}}\right).
\end{equation}
Finally, averaging over $D$ gives
\begin{equation}
\mathbb E[T_f]
\approx \int_0^\infty d\left(\frac{1}{v_c} + \frac{\pi\kappa(1-\zeta)}{2\omega_{\max}}\right)f_D(d)\,\mathrm{d}d,
\end{equation}
which yields the desired \eqref{eq:meanTf_general}.

\bibliographystyle{IEEEtran.bst}
\bibliography{MyRef.bib}

% Generated by IEEEtran.bst, version: 1.14 (2015/08/26)
\begin{thebibliography}{10}
\providecommand{\url}[1]{#1}
\csname url@samestyle\endcsname
\providecommand{\newblock}{\relax}
\providecommand{\bibinfo}[2]{#2}
\providecommand{\BIBentrySTDinterwordspacing}{\spaceskip=0pt\relax}
\providecommand{\BIBentryALTinterwordstretchfactor}{4}
\providecommand{\BIBentryALTinterwordspacing}{\spaceskip=\fontdimen2\font plus
\BIBentryALTinterwordstretchfactor\fontdimen3\font minus
  \fontdimen4\font\relax}
\providecommand{\BIBforeignlanguage}[2]{{%
\expandafter\ifx\csname l@#1\endcsname\relax
\typeout{** WARNING: IEEEtran.bst: No hyphenation pattern has been}%
\typeout{** loaded for the language `#1'. Using the pattern for}%
\typeout{** the default language instead.}%
\else
\language=\csname l@#1\endcsname
\fi
#2}}
\providecommand{\BIBdecl}{\relax}
\BIBdecl

\bibitem{yuan2026}
W.~Yuan, Y.~Cui, J.~Wang, F.~Liu, L.~Zhou, G.~Sun, T.~Xiang, J.~Xu, S.~Jin,
  D.~Niyato, S.~Coleri, S.~Sun, S.~Mao, A.~Jamalipour, D.~I. Kim, M.-S.
  Alouini, and X.~Shen, ``From ground to sky: Architectures, applications, and
  challenges shaping low-altitude wireless networks,'' \emph{arXiv:2506.12308},
  Jan. 2026.

\bibitem{Zhang2023}
T.~Zhang, C.~Chen, Y.~Xu, J.~Loo, and W.~Xu, ``Joint task scheduling and
  multi-{UAV} deployment for aerial computing in emergency communication
  networks,'' \emph{Sci. China Inf. Sci.}, vol.~66, no.~9, p. 192303, Aug.
  2023.

\bibitem{9515586}
T.~Feng, L.~Xie, J.~Yao, and J.~Xu, ``{UAV}-enabled data collection for
  wireless sensor networks with distributed beamforming,'' \emph{IEEE Trans.
  Wireless Commun.}, vol.~21, no.~2, pp. 1347--1361, Feb. 2022.

\bibitem{11311373}
T.~Shi, W.~Wen, P.~Wu, and M.~Xia, ``Vertical heterogeneous networks beyond
  {5G}: {CoMP} coverage enhancement and optimization,'' \emph{IEEE Trans.
  Wireless Commun.}, vol.~25, pp. 9391--9405, 2026.

\bibitem{9861699}
M.~M. Azari, S.~Solanki, S.~Chatzinotas, O.~Kodheli, H.~Sallouha, A.~Colpaert,
  J.~F. Mendoza~Montoya, S.~Pollin, A.~Haqiqatnejad, A.~Mostaani, E.~Lagunas,
  and B.~Ottersten, ``Evolution of non-terrestrial networks from {5G} to {6G}:
  A survey,'' \emph{IEEE Commun. Surv. Tutor.}, vol.~24, no.~4, pp. 2633--2672,
  Aug. 2022.

\bibitem{8673556}
H.~Tabassum, M.~Salehi, and E.~Hossain, ``Fundamentals of mobility-aware
  performance characterization of cellular networks: A tutorial,'' \emph{IEEE
  Commun. Surv. Tutor.}, vol.~21, no.~3, pp. 2288--2308, Mar. 2019.

\bibitem{6477064}
X.~Lin, R.~K. Ganti, P.~J. Fleming, and J.~G. Andrews, ``Towards understanding
  the fundamentals of mobility in cellular networks,'' \emph{IEEE Trans.
  Wireless Commun.}, vol.~12, no.~4, pp. 1686--1698, Apr. 2013.

\bibitem{7110473}
W.~Bao and B.~Liang, ``Stochastic geometric analysis of user mobility in
  heterogeneous wireless networks,'' \emph{IEEE J. Sel. Areas Commun.},
  vol.~33, no.~10, pp. 2212--2225, Oct. 2015.

\bibitem{7006787}
S.~Sadr and R.~S. Adve, ``Handoff rate and coverage analysis in multi-tier
  heterogeneous networks,'' \emph{IEEE Trans. Wireless Commun.}, vol.~14,
  no.~5, pp. 2626--2638, May 2015.

\bibitem{7827020}
R.~Arshad, H.~ElSawy, S.~Sorour, T.~Y. Al-Naffouri, and M.-S. Alouini,
  ``Velocity-aware handover management in two-tier cellular networks,''
  \emph{IEEE Trans. Wireless Commun.}, vol.~16, no.~3, pp. 1851--1867, Mar.
  2017.

\bibitem{10130125}
K.~Tokuyama, T.~Kimura, and N.~Miyoshi, ``Periodic handover skipping in
  cellular networks: Spatially stochastic modeling and analysis,'' \emph{IEEE
  Trans. Wireless Commun.}, vol.~23, no.~1, pp. 91--103, Jan. 2024.

\bibitem{10427200}
B.~Palit and S.~Chakraborty, ``Mobility adapted coverage probability analysis
  in heterogeneous cellular networks,'' in \emph{Proc. 16th Int. Conf.
  COMmunication Systems and NETworkS (COMSNETS)}, 2024, pp. 656--660.

\bibitem{10464354}
C.~Li, N.~Deng, C.~Xing, H.~Wei, N.~Zhao, and D.~Niyato, ``Performance analysis
  of cross-tier interference coordination for mobile users,'' \emph{IEEE Trans.
  Commun.}, vol.~72, no.~8, pp. 5058--5072, Aug. 2024.

\bibitem{camp2002survey}
T.~Camp, J.~Boleng, and V.~Davies, ``A survey of mobility models for ad hoc
  network research,'' \emph{Wireless Commun. Mob. Comput.}, vol.~2, no.~5, pp.
  483--502, Sep. 2002.

\bibitem{Bettstetter2004}
C.~Bettstetter, H.~Hartenstein, and X.~P{\'e}rez-Costa, ``Stochastic properties
  of the random waypoint mobility model,'' \emph{Wireless Netw.}, vol.~10,
  no.~5, pp. 555--567, Sep. 2004.

\bibitem{8671460}
P.~K. Sharma and D.~I. Kim, ``Random {3D} mobile {UAV} networks: Mobility
  modeling and coverage probability,'' \emph{IEEE Trans. Wireless Commun.},
  vol.~18, no.~5, pp. 2527--2538, May 2019.

\bibitem{10777167}
B.~A. Raddwan, I.~Ahmed Al-Baltah, and M.~Ghaleb, ``Environment-aware {3D}
  mobility simulation for the {5G} and {6G} wireless networks,'' in \emph{Proc.
  1st Int. Conf. Emerging Technol. Dependable Internet Things (ICETI)}, 2024,
  pp. 1--8.

\bibitem{8998329}
R.~Amer, W.~Saad, and N.~Marchetti, ``Mobility in the sky: Performance and
  mobility analysis for cellular-connected {UAVs},'' \emph{IEEE Trans.
  Commun.}, vol.~68, no.~5, pp. 3229--3246, May 2020.

\bibitem{10005193}
H.~Sun, C.~Ma, L.~Zhang, J.~Li, X.~Wang, S.~Li, and T.~Q.~S. Quek, ``Coverage
  analysis for cellular-connected random {3D} mobile {UAVs} with directional
  antennas,'' \emph{IEEE Wireless Commun. Lett.}, vol.~12, no.~3, pp. 550--554,
  Mar. 2023.

\bibitem{10125057}
Y.~Guo and H.~Zhang, ``{3D} boundary modeling and handover analysis of aerial
  users in heterogeneous networks,'' \emph{IEEE Trans. Veh. Technol.}, vol.~72,
  no.~10, pp. 13\,523--13\,529, Oct. 2023.

\bibitem{8533634}
M.~{Mozaffari}, A.~T.~Z. {Kasgari}, W.~{Saad}, M.~{Bennis}, and M.~{Debbah},
  ``Beyond {5G} with {UAV}s: Foundations of a {3D} wireless cellular network,''
  \emph{IEEE Trans. Wireless Commun.}, vol.~18, no.~1, pp. 357--372, Jan. 2019.

\bibitem{9151343}
Y.~Li, N.~I. Miridakis, T.~A. Tsiftsis, G.~Yang, and M.~Xia, ``Air-to-air
  communications beyond {5G}: A novel {3D CoMP} transmission scheme,''
  \emph{IEEE Trans. Wireless Commun.}, vol.~19, no.~11, pp. 7324--7338, Nov.
  2020.

\bibitem{9740446}
C.~K. Armeniakos, P.~S. Bithas, and A.~G. Kanatas, ``Finite point processes in
  a truncated octahedron-based {3D} {UAV} network,'' \emph{IEEE Trans. Veh.
  Technol.}, vol.~71, no.~7, pp. 7230--7243, Jul. 2022.

\bibitem{10345607}
K.~Fujieda, T.~Kimura, and T.~Takine, ``Handover analysis in aerial base
  station networks with different altitudes,'' \emph{IEEE Commun. Lett.},
  vol.~28, no.~1, pp. 138--142, Jan. 2024.

\bibitem{9454462}
J.~Huang, J.~Tang, A.~Shojaeifard, Z.~Chen, J.~Hu, D.~K.~C. So, and K.-K. Wong,
  ``Drone mobile networks: Performance analysis under {3D} tractable mobility
  models,'' \emph{IEEE Access}, vol.~9, pp. 90\,555--90\,567, 2021.

\bibitem{moller1989random}
J.~M{\o}ller, ``Random tessellations in {$\mathbb{R}^d$},'' \emph{Adv. Appl.
  Probab.}, vol.~21, no.~1, pp. 37--73, Mar, 1989.

\bibitem{Lautensack_2008}
C.~Lautensack and S.~Zuyev, ``Random {Laguerre} tessellations,'' \emph{Adv.
  Appl. Probab.}, vol.~40, no.~3, pp. 630--650, Sep. 2008.

\bibitem{moller_1992}
J.~Møller, ``Random {Johnson-Mehl} tessellations,'' \emph{Adv. Appl. Probab.},
  vol.~24, no.~4, pp. 814--844, Dec. 1992.

\bibitem{6476039}
Y.~Wan, K.~Namuduri, Y.~Zhou, and S.~Fu, ``A smooth-turn mobility model for
  airborne networks,'' \emph{IEEE Trans. Veh. Technol.}, vol.~62, no.~7, pp.
  3359--3370, Sep. 2013.

\bibitem{8325272}
J.~Xie, Y.~Wan, B.~Wang, S.~Fu, K.~Lu, and J.~H. Kim, ``A comprehensive
  3-dimensional random mobility modeling framework for airborne networks,''
  \emph{IEEE Access}, vol.~6, pp. 22\,849--22\,862, 2018.

\bibitem{shoemake1985animating}
K.~Shoemake, ``Animating rotation with quaternion curves,'' \emph{ACM SIGGRAPH
  Comput. Graph.}, vol.~19, no.~3, pp. 245--254, Jul. 1985.

\bibitem{Mellinger2011}
D.~Mellinger and V.~Kumar, ``Minimum snap trajectory generation and control for
  quadrotors,'' in \emph{Proc. IEEE Int. Conf. Robot. Autom. (ICRA)}, 2011, pp.
  2520--2525.

\bibitem{GarciaMarina2017}
H.~Garcia~de Marina, Y.~A. Kapitanyuk, M.~Bronz, G.~Hattenberger, and M.~Cao,
  ``Guidance algorithm for smooth trajectory tracking of a fixed wing {UAV}
  flying in wind flows,'' in \emph{Proc. IEEE Int. Conf. Robot. Autom. (ICRA)},
  2017, pp. 209--214.

\bibitem{10720695}
Y.~Li, D.~Guo, L.~Luo, and M.~Xia, ``Air-to-ground communications beyond {5G}:
  {CoMP} handoff management in {UAV} network,'' \emph{IEEE Trans. Wireless
  Commun.}, vol.~23, no.~12, pp. 18\,822--18\,837, Dec. 2024.

\bibitem{11126168}
A.~Prado, F.~Mehmeti, and W.~Kellerer, ``Reducing mobility-related signaling
  with network sum throughput maximization in {5G},'' \emph{IEEE Trans. Netw.
  Service Manag.}, vol.~22, no.~6, pp. 6048--6065, Dec. 2025.

\bibitem{chiu2013stochastic}
S.~N. Chiu, D.~Stoyan, W.~S. Kendall, and J.~Mecke, \emph{Stochastic Geometry
  and Its Applications}, 3rd~ed.\hskip 1em plus 0.5em minus 0.4em\relax
  Hoboken, NJ, USA: John Wiley \& Sons, 2013.

\bibitem{Muche_2010}
L.~Muche, ``Contact and chord length distribution functions of the
  {Poisson-Voronoi} tessellation in high dimensions,'' \emph{Adv. Appl.
  Probab.}, vol.~42, no.~1, pp. 48--68, Mar. 2010.

\bibitem{8976426}
Y.~Li, M.~Xia, and S.~Aïssa, ``Coordinated multi-point transmission: A
  {Poisson-Delaunay} triangulation based approach,'' \emph{IEEE Trans. Wireless
  Commun.}, vol.~19, no.~5, pp. 2946--2959, May 2020.

\bibitem{okabe2000spatial}
A.~Okabe, B.~Boots, K.~Sugihara, and S.~N. Chiu, \emph{Spatial Tessellations:
  Concepts and Applications of {Voronoi} Diagrams}, 2nd~ed.\hskip 1em plus
  0.5em minus 0.4em\relax New York, NY, USA: Wiley, 2000.

\bibitem{Alam2006Coverage}
S.~M.~N. Alam and Z.~J. Haas, ``Coverage and connectivity in three-dimensional
  networks,'' in \emph{Proc. 12th Annu. Int. Conf. Mobile Comput. Netw.
  (MobiCom)}, Los Angeles, CA, USA, 2006, pp. 346--357.

\bibitem{7058418}
Z.~Pan and Q.~Zhu, ``Modeling and analysis of coverage in {3-D} cellular
  networks,'' \emph{IEEE Commun. Lett.}, vol.~19, no.~5, pp. 831--834, May
  2015.

\bibitem{schneider2008stochastic}
R.~Schneider and W.~Weil, \emph{Stochastic and Integral Geometry}, ser.
  Probability and Its Applications.\hskip 1em plus 0.5em minus 0.4em\relax
  Berlin, Heidelberg: Springer, 2008.

\bibitem{105566}
L.~M. Cruz-Orive, ``Stereology: A historical survey,'' \emph{Image Anal.
  Stereol.}, vol.~36, no.~3, pp. 153--177, Jan. 2017.

\bibitem{Gallager2013}
R.~G. Gallager, \emph{Discrete Stochastic Processes}.\hskip 1em plus 0.5em
  minus 0.4em\relax Cambridge University Press, 2013.

\bibitem{WeissteinSpherePointPicking}
\BIBentryALTinterwordspacing
E.~W. Weisstein, ``Sphere point picking,'' MathWorld---A Wolfram Resource,
  accessed: 2026-03-29. [Online]. Available:
  \url{https://mathworld.wolfram.com/SpherePointPicking.html}
\BIBentrySTDinterwordspacing

\end{thebibliography}

\begin{IEEEbiography}
	[{\includegraphics[width=1in,height=1.25in, clip, keepaspectratio]{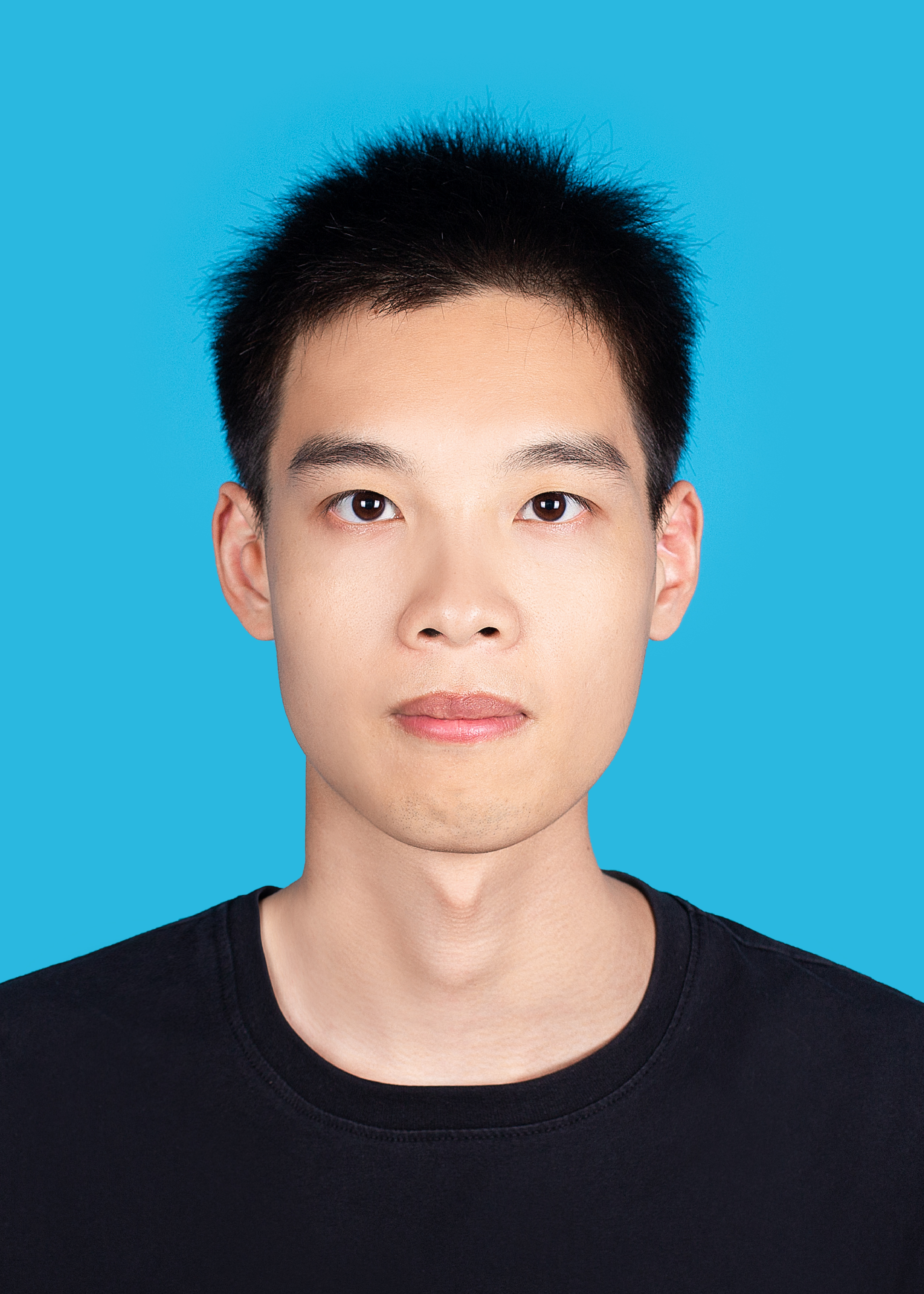}}]{Tian Shi} received the B.Sc. degree in information and computing science in 2019 and the M.Sc. degree in mathematics in 2023, both from Guilin University of Electronic Technology, Guilin, China. He is currently pursuing a Ph.D. degree in information and communication engineering at Sun Yat-sen University, Guangzhou, China. His research interests include stochastic geometry and cooperative UAV communications.
\end{IEEEbiography}

\begin{IEEEbiography}
	[{\includegraphics[width=1in, height=1.25in, clip, keepaspectratio]{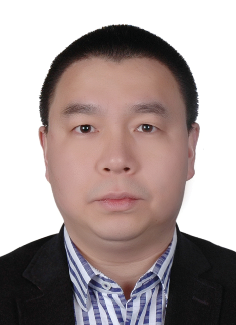}}]{Minghua Xia} (Senior Member, IEEE) received the Ph.D. degree in Telecommunications and Information Systems from Sun Yat-sen University, Guangzhou, China, in 2007.
	
	From 2007 to 2009, he was with the Electronics and Telecommunications Research Institute (ETRI) of South Korea, Beijing R\&D Center, Beijing, China, where he worked as a member and then as a senior member of the engineering staff. From 2010 to 2014, he was in sequence with The University of Hong Kong, Hong Kong, China; King Abdullah University of Science and Technology, Jeddah, Saudi Arabia; and the Institut National de la Recherche Scientifique (INRS), University of Quebec, Montreal, Canada, as a Postdoctoral Fellow. Since 2015, he has been a Professor at Sun Yat-sen University. Since 2019, he has also been an Adjunct Professor with the Southern Marine Science and Engineering Guangdong Laboratory (Zhuhai). His research interests are in the general areas of wireless communications and signal processing.
\end{IEEEbiography}

\vfill
\end{document}